%% file: YongSilva22.tex
\documentclass[review]{elsarticle}

\usepackage{lineno,hyperref}

\journal{Combustion and Flame}

\usepackage{graphicx}
\usepackage{fancyhdr}

\usepackage{url}
\usepackage{tabto}
\input{math.tex}

\usepackage[toc,page]{appendix}

\usepackage{amsmath}
\usepackage{bm}
\usepackage{mathrsfs} 
\usepackage{amssymb}

\usepackage{dashrule}

\usepackage{graphics}
\usepackage{setspace}
\usepackage{multirow}

\usepackage[english]{babel}

\usepackage{xcolor}

\usepackage[utf8]{inputenc}
\usepackage[T1]{fontenc}
\usepackage{array}
\usepackage{booktabs}
\usepackage{csquotes}
\usepackage{algorithm}
\usepackage{algpseudocode}

\usepackage{scrextend}

\usepackage{tabularx}
\newcolumntype{Y}{>{\centering\arraybackslash}X}

\usepackage{scalerel,amssymb}

\usepackage{mathtools}
\usepackage{subcaption}

\newcommand{\cmnt}[1]{\textcolor{black}{#1}}
\newcommand{\criterionName}{{\Phi}}
\newcommand{\criterionNameText}{{Phi}}

\newlength{\mywidth}
\newlength{\figwidth}
\biboptions{sort&compress} 
\begin{document}
\include{Chapters/YongSilva22_Main}
\appendix

\include{Chapters/YongSilva22_Appendix}
\end{document}

%% file: math.tex
\def\t2d#1#2{{d^2 #1\over d #2^2}}                

\def\p2d#1#2{{\partial^2 #1\over\partial #2^2}}

\newenvironment{matrix1}{ \left( \begin{array}{c} } { \end{array} \right) }
\newenvironment{matrix2}{ \left( \begin{array}{cc} } { \end{array} \right) }
\newenvironment{matrix3}{ \left( \begin{array}{ccc} } { \end{array} \right) }

\renewcommand{\Re}{\, \mbox{Re}} 

%% file: Chapters/YongSilva22_Main.tex
\begin{frontmatter}

\title{Categorization of Thermoacoustic Modes in an Ideal Resonator with Phasor Diagrams} 


\author{Kah Joon Yong\corref{mycorrespondingauthor}}
\cortext[mycorrespondingauthor]{Corresponding author}
\ead{yong@tfd.mw.tum.de}

\author{Camilo F. Silva}

\author{Guillaume J. J. Fournier}

\author{Wolfgang Polifke}

\address{TUM School of Engineering and Design,
Department of Engineering Physics and Computation, Technical University of Munich, D-85747 Garching, Germany}

\begin{abstract}
A recent study (\textit{Yong, Silva, and Polifke, Combust. Flame 228 (2021)}) proposed the use of phasor diagrams to categorize marginally stable modes in an ideal resonator with a compact, velocity-sensitive flame. Modes with velocity phasors that reverse direction across the flame were categorized as ITA modes. 
The present study extends this concept to growing and decaying modes. In other words, with the method proposed, it is possible to distinguish whether a given thermoacoustic mode -- regardless of its stability -- should be categorized as acoustic or ITA. The method proposed does not rely on any parametric sweep, but on the angle relating the velocity phasors across the flame. This method of categorization reveals distinct regions in the complex plane where acoustic and ITA eigenfrequencies are localized. Additionally, we analyze the medium oscillation at the flame location to construct a physically intuitive understanding of the proposed categorization method.

\end{abstract}


\end{frontmatter}

\nolinenumbers

\section{Introduction}

Thermoacoustic combustion instabilities (TCI) have been the subject of investigation over the past several decades due to their detrimental effects on combustion systems. 
Recently, the increased demand for cleaner, more efficient and more flexible combustion technologies has driven the need for lean premixed combustors, which are deemed more susceptible to TCI \cite{Cande02,LieuwMcMan03,Poins17}. 
The onset of TCI is attributed to a positive feedback coupling between a fluctuating flame and acoustic perturbations \cite{Cande02,PoinsVeyna05,Dowli95, LieuwYang05}. Acoustic waves are generated by the unsteady volumetric expansion in the reactive region. According to the established understanding, these acoustic waves propagate upstream or downstream until reaching the acoustic boundaries, where they are reflected back towards the flame. At the flame, the reflected waves may perturb the flow variables associated with the flame such as the equivalence ratio, the swirl number, and especially the velocity of the incoming premixture, which in turn perturb the heat release rate, thus closing the feedback loop. If the heat release rate fluctuations are in-phase with the pressure fluctuations (fully or partially), i.e. if heat is given to the medium at the moment of greatest condensation, energy is transferred from the flame to the acoustic field \cite{Rayle78}. Provided that acoustic dissipation is low, self-excited oscillations may ensue. \cmnt{
However, in the feedback loop mechanism described above, the flame may be insensitive to acoustic fluctuations, i.e. a passive flame that only induces a temperature discontinuity. In this case, the reflected acoustic waves result in a standing wave mode -- `\emph{pure} acoustic mode'. Such mode is either marginally stable or damped, if dissipation and losses are taken into account.}

\begin{figure}[ht]
\centering
\includegraphics[width=0.7\textwidth]{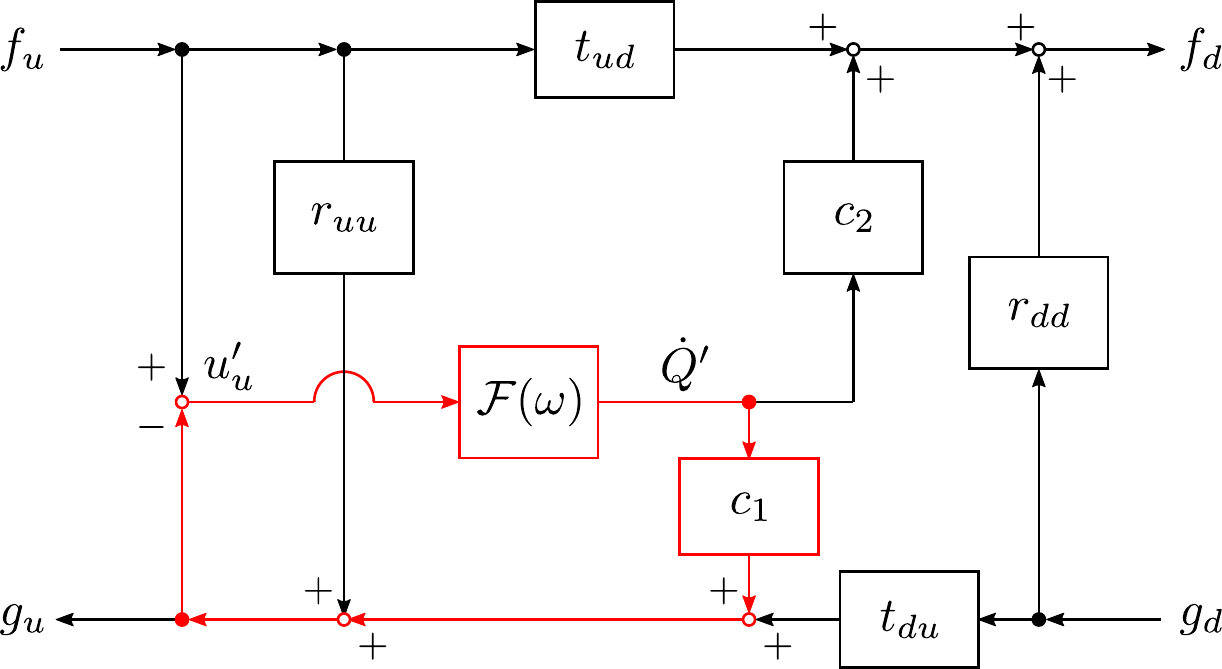}
\caption{The signal flow chart visualizes the scattering pathways of the characteristic wave amplitudes $f_{[\ ]},g_{[\ ]}$ at the discontinuity due to the presence of a velocity sensitive flame. The indexes $[\ ]_u$ and $[\ ]_d$ indicate the location at the immediate \underline{u}pstream and \underline{d}ownstream of the flame.  $r_{uu},t_{ud},t_{du}, r_{dd}$ are scattering matrix elements that describes the acoustic wave scattering \cmnt{due to the an impedance change across the flame interface (see \cite{EmmerBombe15} for the exact definitions)}. \cmnt{$c_1, c_2$ are the gain factors that correspond to the production of acoustic waves traveling upstream and downstream $g_u,f_d$ due to the fluctuating heat release rate $\dot{Q}'$.}
The red lines highlight the ITA feedback mechanism: the upstream velocity perturbation $u'_u$ triggers a flame response $\dot{Q}'$ described by the flame transfer function $\mathcal{F}(\omega)$. The flame response facilitates the production of an acoustic wave travelling upstream $g_u$, which in turn mediates $u'_u$, closing the feedback loop. }
\label{fig:ITALoop}
\end{figure}

This description of flame-acoustic feedback was shown to be incomplete, as the `intrinsic thermoacoustic (ITA) feedback loop' was discovered \cite{HoeijKorni14,EmmerBombe15}. This ITA loop may be visualized in a signal flow chart as in Fig. \ref{fig:ITALoop} \cmnt{(red pathway)}.
\cmnt{This feedback loop describes a more immediate route for the flame-acoustic perturbation to occur without involving the boundary reflections: the upstream velocity perturbation $u'_u$ triggers a flame response $\dot{Q}'$, which in turn generates an acoustic wave $g_u$ that travels upstream to perturb the upstream velocity, thereby closing the loop \cite{HoeijKorni14,EmmerBombe15}. Thermoacoustic modes generated solely through this feedback loop are known as `\emph{pure} ITA modes'. We may deduce from Fig. \ref{fig:ITALoop} that all modes observed in a system with non-vanishing boundary reflections and a fluctuating flame are the result of the interplay between both the acoustic and ITA feedback loops. Indeed, Emmert et al. \cite{EmmerBombe17} introduced the notion that a thermoacoustic mode in an echoic chamber can be predominantly associated either with the ITA loop -- `ITA modes' -- or with the acoustic loop -- `acoustic modes', after demonstrating that the number of modes in a closed-open combustor exceeds the number of acoustic modes}. The results of other numerical studies, which examined the changes in frequency and growth rate as the boundary reflection coefficients gradually increased from zero \cite{HoeijKorni14,SilvaEmmer15}, are consistent with this assertion. 

The ability to distinguish ITA modes from acoustic modes becomes important, when one tries to understand if certain behaviour observed either in simulations or in experiments are inherent to ITA modes. One of the most discussed behaviors, which was attributed to ITA modes in the studies \cite{SilvaMerk17, EmmerBombe17}, is mode destabilization upon boundary damping. This phenomenon was also confirmed in an experiment \cite{XuZheng20}, where an acoustic liner at the end of the combustor suppressed instabilities associated with acoustic modes, but promoted those associated with ITA modes. Furthermore, eigenfrequencies of ITA modes are thought to be less sensitive to changes in combustor length than acoustic modes \cite{SilvaMerk17,MurugSinga18}. This conclusion was supported by the fact that the intrinsic feedback loop of the flame does not require acoustic waves traveling through the entire combustor (see above). 

A variety of ad hoc methods were proposed to achieve the goal of mode categorization. 
Emmert et al \cite{EmmerBombe17} introduced a coupling parameter $\mu$ to modulate the coupling between the acoustic and ITA feedback loops in a thermoacoustic system, \cmnt{i.e. the coupling between the fluctuating heat release rate $\dot{Q}'$ and the characteristic wave amplitudes $f,g$.} 
As $\mu$  was gradually reduced from unity to 0, the existing eigenmodes either tracked to a `pure ITA mode'-- modes found exclusively in an anechoic environment -- or to a `pure acoustic mode' -- modes found in a system with a passive flame (flame that is insensitive to acoustic fluctuations). These modes were thus categorized as ITA and acoustic modes, respectively. 
In the same vein, other authors \cite{MukheShrir17,SilvaYong18a, OrchiSilva20,HosseKorni18a} performed the sweep of parameters such as the flame gain or the reflection coefficients.
Alternatively, proximity to a pure ITA or pure acoustic mode is used to identify an ITA or acoustic mode \cite{SilvaEmmer15,SogarSchmi19, BuschMensa20,FournHaeri20a} . Other studies, including \cite{AlbayStein17a, XuZheng20}, simply made use of the established characteristics of pure ITA modes, say, independence of eigenfrequency from combustor length or eigenfrequency correspondence with a characteristic time delay of the flame response, to identify ITA modes. 

Aside from being impractical in situations where parametric variation is difficult or expensive, these methods suffer from multiple limitations. For instance, methods that rely on the closeness of a mode eigenfrequency to the corresponding frequency of pure acoustic or ITA modes, are only applicable to a limited number of modes. A parameter sweep often entails a stark variation of system parameters, which modifies the nature of the modes of interest. Indeed,
Mukherjee et al. \cite{MukheShrir19} observed that the pressure profile of the ITA modes and the acoustic modes categorized by the tracking of the flame gain become indistinguishable from each other for a moderately large gain. As an acoustic mode `approaches a nearby ITA mode, it stops behaving like an acoustic mode'. The concept of modes switching identities under parametric variation was asserted by Hosseini et al. \cite{HosseKorni18a}, as a complex interplay between ITA and acoustic modes was demonstrated under variation of parameters such as time delay, reflection coefficients, or temperature ratio across the flame. This assertion was reinforced by the results in \cite{SilvaYong18a, SogarSchmi19}. The lack of a physically based categorization method has led to the claim that mode identities do not correlate with specific mode behaviors. As such, Mukherjee et al. \cite{MukheShrir19} speak of `born acoustic' or `born ITA' modes, which are modes that tracked to pure acoustic or pure ITA modes at the limit of small gain and have no correlation to the established characteristics typical for a pure acoustic or pure ITA mode. Similarly, Orchini et al. \cite{OrchiSilva20} classified the eigenmodes into modes of `acoustic origin' and `ITA origin', while highlighting the shared mode shape features between them in the proximity of an exceptional point. 

Yong et al. \cite{YongSilva21} proposed a method that does not involve a parameter sweep to categorize marginally stable modes in an ideal cavity. Phasors and phasor diagrams were used to represent the acoustic variables, i.e. the characteristic wave amplitudes and the fluctuations in pressure, velocity and heat release rate. The phasor diagram in an anechoic environment reveals that pure ITA modes exhibit velocity phasors that are perfectly out of phase across the flame, which implies a change in sign of the pressure gradient across the flame. Interestingly, the same phasor characteristic is applicable to marginally stable modes in an ideal resonator, providing that the respective flame is sufficiently strong in gain and acts in the opposite direction of the upstream velocity. Due to the analogous phasor characteristic to pure ITA modes, these marginally stable modes are subsequently categorized as ITA modes. \cmnt{This method of ITA mode categorization is coherent with the previous works that investigated pure ITA modes \cite{HoeijKorni14,EmmerBombe15,MukheShrir17}, which concluded that ITA modes may arise when the FTF phase is equal to an odd multiple of $\pi$ -- the `$\pi$-criterion'.}

Unlike the previously proposed methods, the phasor based categorization proposed by Yong et al. is physically motivated, not limited to a specific flame model such as the $n-\tau$ model, applicable in a variety of combustor setups, and also readily implementable in experimental settings. 
However, that proposed criterion has its limitation as Yong et al.~\cite{YongSilva21} only investigated the marginally stable modes in an ideal resonator. 
In this work, we extend the scope of the phasor analysis to growing or decaying thermoacoustic modes, which allows the generalization of the phasor criterion.

This paper is structured as follows: The next section introduces phasor diagrams for wave propagation and thermoacoustic coupling across a compact heat source for non-zero growth rates.  The differences to the phasor diagram of a marginally stable mode are highlighted. In the process, the influence of the non-zero growth rates on the flame transfer function (FTF) in terms of magnitude and phase are discussed. The second section serves as a preparatory stage for the core findings in the following sections. Here, the categorization of marginally stable thermoacoustic modes that relies on the direction reversal of the velocity phasors across the flame is recapitulated. Then, two limiting cases -- highly decaying and highly unstable modes -- are analyzed using the previous criterion. The third section inspects the phasor diagrams of a large variety of moderately  unstable and decaying modes. Based on their distribution in a stability map, modes with partially aligned or partially anti-aligned velocity phasors across the flame are categorized as acoustic or ITA, respectively. It is demonstrated that a continuous increase or decrease in growth rates necessarily leads to a continuous transition of marginally stable acoustic modes into ITA modes at infinite growth and decay rates. In the subsequent section, the relative orientation of the velocity phasors is quantified in terms of the scalar product and visualized in contour maps that reveal the regions of acoustic and ITA modes. In the contour maps, an alternating pattern of decreasing and increasing size of the acoustic regions along the real axis is observed. This pattern is shown to directly correspond to the flame position. To complete the section, the transitional magnitude and phase of the FTF are derived. Finally, a physical interpretation is given to the categorization criterion, where the velocity phasors at the flame are interpreted as the physical motion of the flame medium. Acoustic modes and ITA modes are described as modes that predominantly exhibit a back-and-forth versus an inwards-and-outwards oscillation at the flame, respectively. The paper ends with summary, conclusions and outlook. 

Note that this work is a follow-up study of \cite{YongSilva21}. Thus, we highly recommend the readers to read the previous paper, to gain a better understanding on the construction of phasors and phasor diagrams, as well as the notations and keywords used.

\section{Phasor diagrams of characteristic wave amplitudes for non-zero growth rates}

\begin{figure}[ht]
\centering
\includegraphics[width=0.5\textwidth]{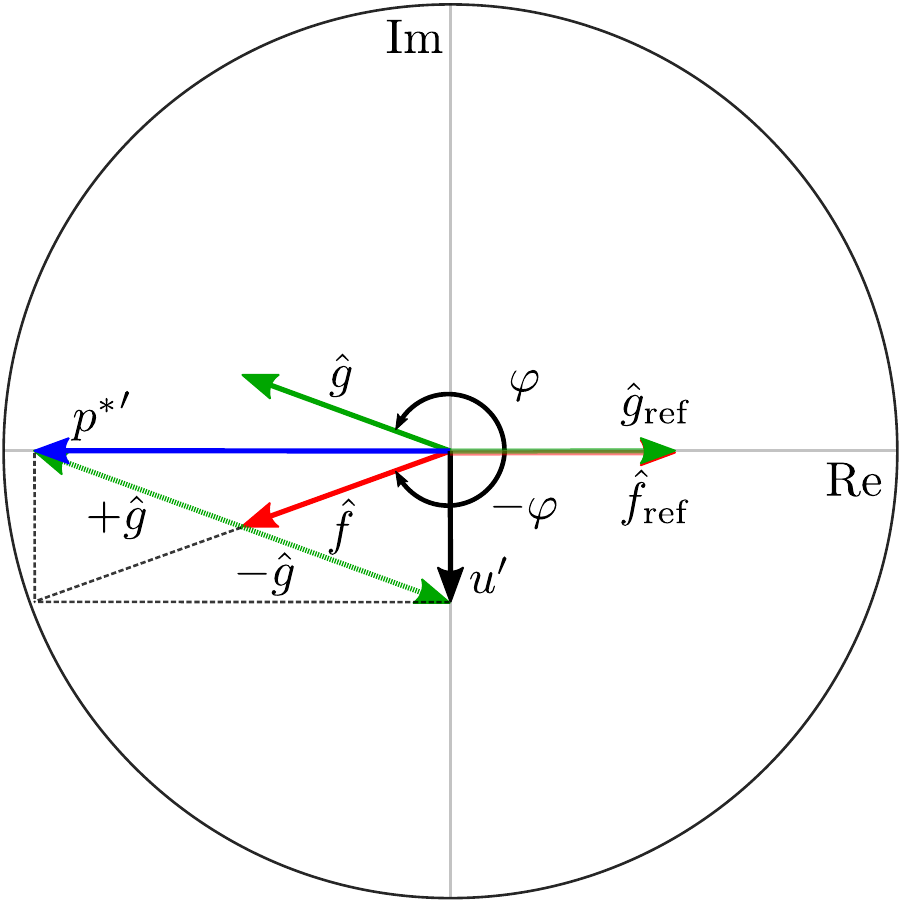}
\caption{A 2D phasor diagram depicting the evolution of the phasors of the CWAs $\hat{f},\hat{g}$ in the $+x$ direction. The phase advance of the $\hat{f},\hat{g}$ relative to the reference position $x=0$ is given by $\varphi=kx$ in the clockwise- and counterclockwise directions, respectively, according to the the factor $e^{\mp ikx}$ given in Eq. \eqref{eq:cwa}. It is important to note that the time harmonic factor $e^{i\omega t}$ is factored out in a phasors representation. The phasor of the primitive variables ${p^*}',u'$ are obtained through addition and subtraction between the $\hat{f},\hat{g}$ phasors, according to Eq. \eqref{eq:fg2pu}.}
\label{fig:phasorBasic}
\end{figure}

In this section, we explore the spatial dependency of the characteristic wave amplitudes of modes with non-zero growth, and produce the corresponding phasor diagrams. 
The propagation of 1D plane waves may be expressed in terms of characteristic waves $f$ and $g$ 
\begin{equation}
f(x,t) = \hat{f}(x)e^{i\omega t}= \hat{f}(x_\text{ref}) e^{i\omega t-ik x}, \quad g(x,t) = \hat{g}(x)e^{i\omega t} = \hat{g}(x_\text{ref}) e^{i\omega t+ik x} 
\end{equation}
where 
\begin{equation}
    \hat{f}(x) = \hat{f}(x_\text{ref}) e^{-ik x}, \quad \hat{g}(x) = \hat{g}(x_\text{ref}) e^{ik x} \label{eq:cwa}
\end{equation}
describes the characteristic wave amplitudes (CWA), which propagate at the mean speed of sound $\overline{c}$ in $\pm x$ direction, respectively; $k=\omega/\overline{c}$ is the wave number for a medium at rest; $\omega$ is the eigenfrequency; $x_\text{ref}$ indicates a reference location. As shown in \cite{YongSilva21}, for real-valued frequencies $\omega\in \mathbb{R}$, the CWAs in Eq. \eqref{eq:cwa} describe a respective $f$ and $g$ phasors rotation in the clockwise and counterclockwise direction as $x$ increases, with an angle of rotation $\varphi=k x$. Figure \ref{fig:phasorBasic} illustrates the 2D phasor diagram at an arbitrary location $x$ in a cavity with the reflection coefficient at $x_\textnormal{ref}$ being
\begin{equation}
    R_\text{ref}=\frac{\hat{f}(x_\text{ref})}{\hat{g}(x_\text{ref})}=1. \label{eq:idealBC}
\end{equation}
The phasors of the primitive acoustic variables, i.e. fluctuations of normalized pressure $p'^{*}=p'/(\overline{\rho}\overline{ c})$ and velocity $u'$ are obtained by adding and subtracting $f,g$ phasors, according to the relation
\begin{equation}
{p^{*}}'=\frac{p'}{\overline{\rho c}} = \hat{f}+\hat{g}; \quad u'=\hat{f}-\hat{g}, \label{eq:fg2pu}
\end{equation}
where $\overline{\rho}$ and $\overline{\rho}\overline{ c }$ represent mean density and specific impedance, respectively. 
For brevity, the apostrophe $[\ ]'$, which indicates a fluctuation, and the hat $[\hat{\ }]$, which indicates a Fourier transformed variable, are omitted in all following figures as well as texts and equations. The overline $\overline{[\ ]}$ remains to indicate a mean value.  
A series of 2D phasor diagrams at different locations may be combined to generate a 3D phasor plot to depict the complete phasor evolution along the length of the cavity $x\in[0,L]$, as shown in Fig. 3 in \cite{YongSilva21}.
The readers are invited to verify that the constraints of ideal boundaries and real-valued $\omega$ always result in a mirror symmetrical rotation of $f,g$. As a consequence, $p^*$ and $u$ are always in quadrature throughout the resonator. 

\begin{figure}[ht]
\centering
\includegraphics[width=0.5\textwidth]{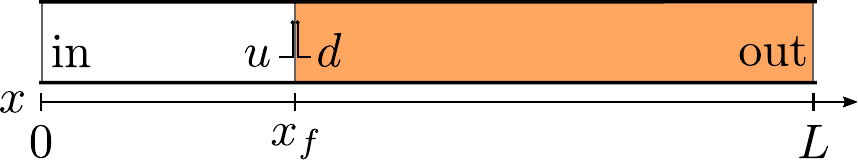}
\caption{Schematic representation of a simple quasi-1D cavity with a velocity-sensitive flame located at $x_f$. The locations of inlet, outlet, flame upstream and flame downstream are indicated by `in', `out', `$u$', and `$d$' respectively. These labels are used as indices for the variables throughout this paper to point to the given specific locations. The inlet at $x = 0$ and the outlet at $x = L$ are ideally closed ($u_\text{in} = 0$) and open ($p_\text{out} = 0$), respectively. The `cold' and `hot' sections are colored white and orange, with constant mean temperatures $T_u$ and $T_d$, respectively. }
\label{fig:DuctSchematics}
\end{figure}
\begin{figure}[ht]
\centering
  \centering
  \includegraphics[width=0.7\linewidth]{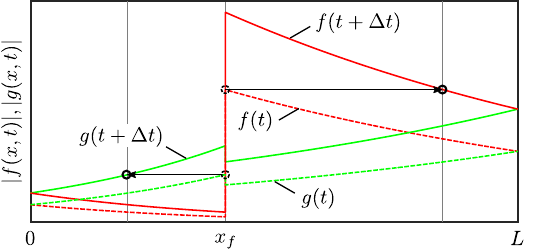}
\caption{The amplitude of the characteristic waves $|f|$ (red) and $|g|$ (green) as a function of $x$ at different time delay $\Delta t$ for a growing thermoacoustic mode $\sigma>0$. Without essential loss of generality, the flame position is chosen to be the reference position, $x_\textnormal{ref}=x_f$. The dashed and solid lines indicate the profiles at initial time and delayed time respectively. The arrows show the direction of wave propagation. In general, $f$ and $g$ grow with time due to energy production at the flame. As time progresses, they propagate without loss or growth across the ideal resonator, resulting in the decaying profile in the direction of propagation.} \label{fig:fgDist}
\end{figure}

In the case of net generation or loss of acoustic energy at the compact elements (flame, area change, damper, ...) or at the system terminations, $\omega$ may be complex valued,
\begin{equation}
\omega = \omega_r + i\omega_i = \omega_r - i\sigma, \label{eq:omega}
\end{equation}
where $\omega_r$ is the angular frequency and $\sigma$ is the growth rate, with $\sigma>0$ describing a growth of acoustic variables $f,g$ as well as $p,u$ in time. Recalling \eqref{eq:cwa} and the dependence of wave number on frequency $k=\omega/\overline{c}$ , it is evident that the $f$ phasor amplitude decreases, while the $g$ phasor amplitude increases with $x$, in the case of $\sigma>0$. It appears counter-intuitive, but a growth in the time domain corresponds to a decay in space. This characteristic is visualized in Fig. \ref{fig:fgDist}. From the flame location $x_f$, $f$ and $g$ waves propagate upstream and downstream at constant amplitude, respectively. As time progresses, stronger characteristic waves are being produced at $x_f$ due to the positive growth. As a result, the wave profiles appear to be decaying in the direction of wave propagation. 

\subsection{Flame-acoustic coupling}
Growing or decaying thermoacoustic modes are the result of imbalanced acoustic gain and loss.
In this work, we consider an active flame, i.e. a flame with unsteady heat release, the sole mechanism for the gain or loss of acoustic energy in an ideal \cmnt{closed-open} resonator as depicted in Fig. \ref{fig:DuctSchematics}. At low Mach numbers, a compact premixed flame introduces a discontinuity in the acoustic field in two ways: first, the increase in mean temperature across the flame changes the specific impedance, thus also the partial transmission and reflection of acoustic waves; second, the unsteady heat release rate modifies the relation of velocity perturbation (magnitude and phase) between locations just up- and downstream of the flame. These effects are summarized in the coupling relations for acoustic variables across the flame -- the linearized `Rankine-Hugoniot' (RH) relations \cite{Polif07,KopitPolif08,Kelle95,StrobBombe16}. For low Mach number flow ($\overline{u}_u \ll \overline{c}_u$), the RH relations reduce to 
\begin{align}
p_d^*=\xi p_u^*, \quad
u_d = u_u +\underbrace{\theta \overline{u}_u \frac{\dot{Q}}{\overline{\dot{Q}}} }_{u_q}, \label{eq:RH}
\end{align}
where the indices $[\ ]_u$ and $[\ ]_d$ indicate the flame upstream and downstream positions,  $\xi=(\overline{\rho}_u \overline{c}_u)/(\overline{\rho}_d \overline{c}_d)$ is the ratio of specific impedance, $\theta = \overline{T}_d/\overline{T}_u-1$ is the relative temperature increment,  and $u_q$ is the displacement rate contribution due to heat release rate fluctuation $\dot{Q}$\footnote{Without essential loss of generality, $u_q$ is used interchangeably to represent $\dot{Q}$ in this work. These quantities are proportional and in-phase.}.
As summarized in the Rayleigh index (RI)\cite{Rayle78,PoinsVeyna05,LieuwYang05}, acoustic energy will be produced by the flame, if the heat release rate fluctuations are (fully or partially) in-phase with the pressure perturbations, 
\begin{equation}
    \textnormal{RI} \equiv \oint \ p\dot{Q}\ \textnormal{d}t > 0. \label{eq:RI}
\end{equation}
In an ideal resonator, where acoustic energy is not lost through the boundaries or flow entropies, a positive Rayleigh index at the flame corresponds directly to the onset of instabilities, which is represented by a positive growth rate $\sigma>0$. 

\begin{figure}[ht]
\centering
\begin{subfigure}{.45\textwidth}
    \centering
    \includegraphics[width=\linewidth]{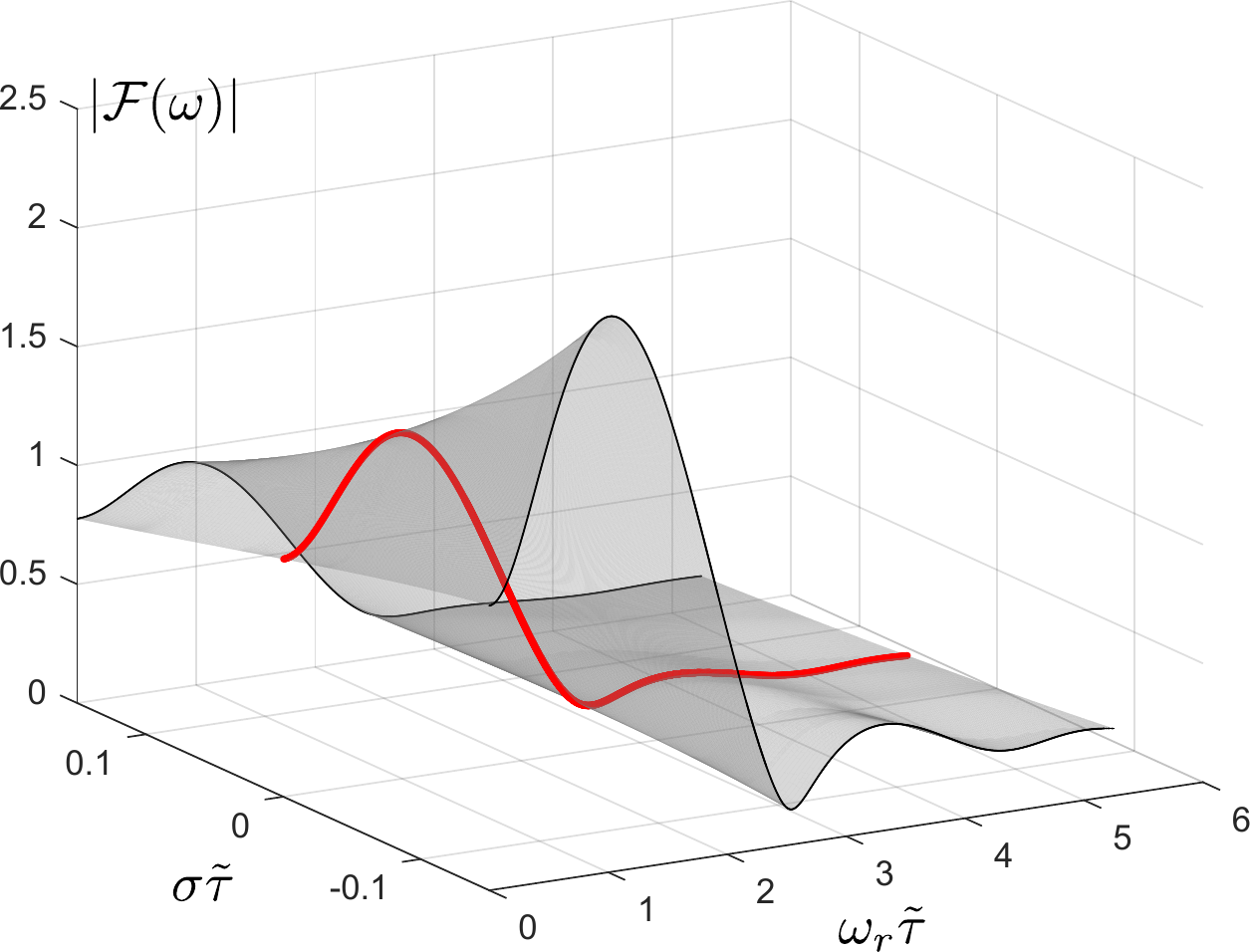}%
    \caption{}
\end{subfigure}%
\begin{subfigure}{.45\textwidth}
    \centering
    \includegraphics[width=\linewidth]{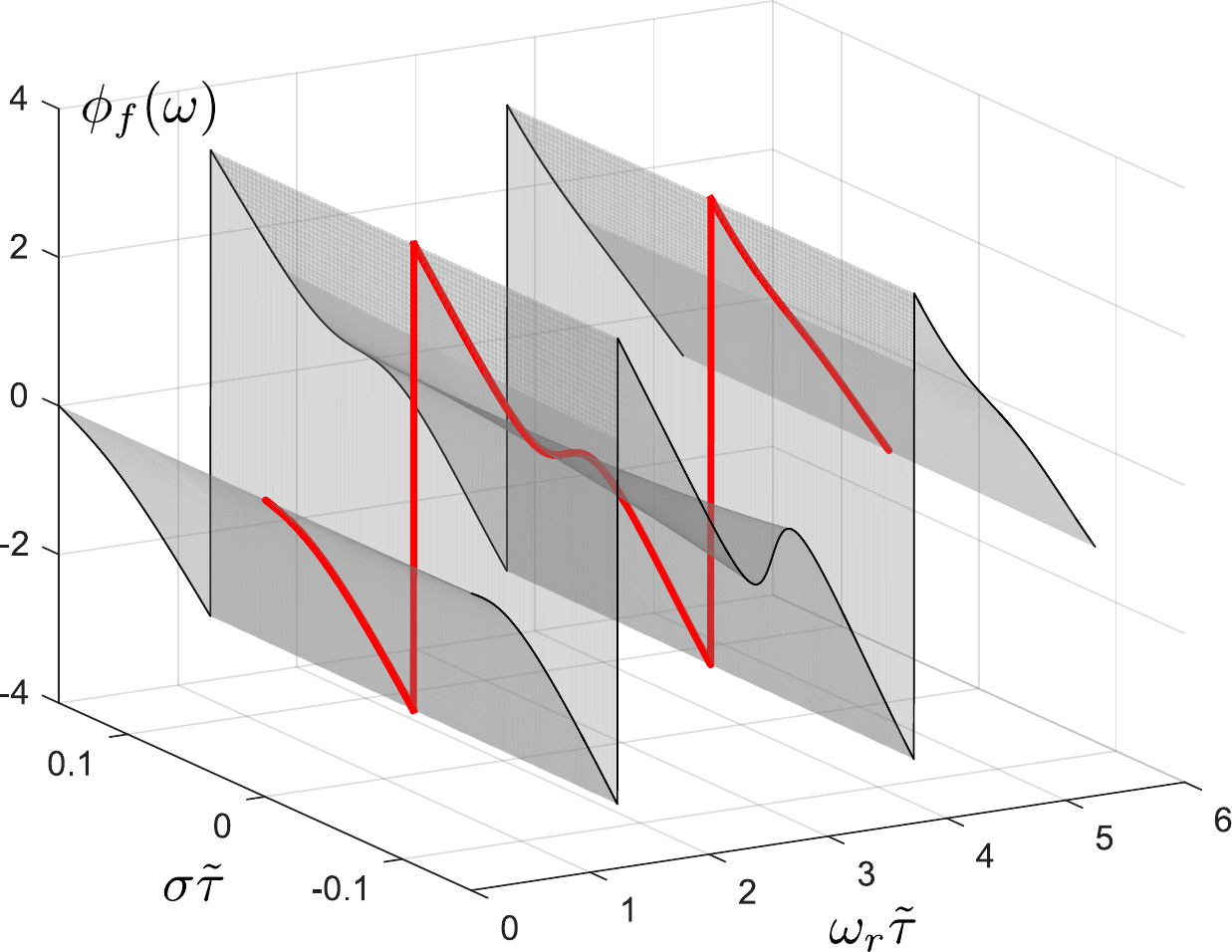}%
    \caption{}
\end{subfigure}
\caption{Surface plot of the (a) FTF magnitude $|\mathcal{F}(\omega)|$ and (b) phase $\phi_f(\omega)$ of a DTD model with $h_k$ given in Tab. \ref{tab:DTDPar},  $\tilde{\tau}=10\Delta t$, with respect to the oscillation frequency $\omega_r$ and growth rate $\sigma$. The inscribed red curves along the real frequency axis mark the FFR. An increase in $\sigma$ reduces the magnitude of the FTF, but has a minimal impact on its phase.} \label{fig:DTD}
\end{figure}

\cmnt{The heat release rate of a compact premixed flame is velocity sensitive \cite{FleifAnnas96,DucruDurox00,LieuwYang05,SilvaEmmer15}}. As such, the dependency of the heat release rate fluctuation $\dot{Q}$ on the velocity perturbation on the flame upstream $u_u$ may be characterized by a frequency dependent \emph{flame transfer function} (FTF) $\mathcal{F}(\omega)$ as given in Eq. \eqref{eq:FTF}.
\begin{equation}
    \frac{\dot{Q}(\omega)}{\overline{\dot{Q}}} = \mathcal{F}(\omega)\frac{{u}_u(\omega)}{\overline{{u}}_u} \label{eq:FTF}
\end{equation}
In general, the FTF may be written as
\begin{equation}
    \mathcal{F}(\omega)=|\mathcal{F}(\omega)|e^{i\phi_f(\omega)} \label{eq:FTF_general}.
\end{equation}
where $|\mathcal{F}(\omega)|$ is the magnitude, and $\phi_f(\omega)$ is the phase of the FTF relative to the upstream velocity perturbation. 
For stability analysis, the FTF needs to be evaluated over all complex frequencies $\omega\in \mathbb{C}$. It should be distinguished from the \emph{flame frequency response} (FFR) $F(\omega_r)$, which describes the flame response at real valued frequencies \footnote{\cmnt{Note that in the literature there is often no strict distinction between the FTF and the FFR. We find it helpful to distinguish between them here, as the FTF magnitude incorporates a non-zero growth rate (see above), which goes against the intuitive understanding of flame strength, i.e. FFR gain. }}
\begin{equation}
    F = \mathcal{F}(\omega_r). \label{eq:FRF}
\end{equation}
This distinction is important, as for a given frequency $\omega_r$ and growth rate $\sigma \neq 0$, the magnitude of the FTF, does not equal the \emph{gain} of the FFR. 
To illustrate the dependency of the FTF on $\sigma$, we consider, by way of example, the simple
$n-\tau $ model \cite{Crocc51}
\begin{equation}
    \mathcal{F}(\omega) = n e^{-i\omega\tau} = ne^{-\sigma\tau}\ e^{-i\omega_r\tau}. \label{eq:FTF_ntau}
\end{equation}
Comparing Eqs. \eqref{eq:FTF_general}and \eqref{eq:FTF_ntau} gives the FTF magnitude and phase
\begin{equation}
    |\mathcal{F}(\omega)| =  n e^{-\sigma\tau}, \quad \phi_f(\omega)=-\omega_r\tau,
\end{equation}
where $n$ represents the gain of the corresponding FFR. In this model, $\sigma$ contributes purely to the FTF magnitude, while the oscillation frequency $\omega_r$ contributes to the phase. 
In a more realistic flame model that exhibits excess gain and a low-pass behaviour, such as the distributed time delay (DTD) model \cite{Polif20}
\begin{equation}
    \mathcal{F}(\omega) = \sum_{k=0}^{+\infty} h_ke^{-i\omega k\Delta t} = \sum_{k=0}^{+\infty} h_ke^{-\sigma k\Delta t}\ e^{-i\omega_r k\Delta t}, \label{eq:FTF_DTD}
\end{equation}
with the impulse response coefficients $h_k$, and the sampling time step $\Delta t$, both the FTF magnitude and phase are influenced by the growth rate and frequency. However, due to the exponential scaling of $e^{-\sigma k\Delta t}$, one might expect that non-zero growth rates have a greater impact on the FTF magnitude than on the phase. From Eq. \eqref{eq:FTF_DTD}, it is not difficult to see that $\sigma\gg0$ will result in a reduction of the corresponding weights of the impulse response $h_k$, i.e. $e^{-\sigma k\Delta t}\ll1$, and thus an overall reduction of the FTF magnitude. As an example, the FTF magnitude and phase of a DTD modeled flame with eight coefficients $h_k$ (c.f. Tab. \ref{tab:DTDPar}) are depicted in Fig. \ref{fig:DTD}.
Essentially, we observe that an increase in growth rate generally results in the reduction of $|\mathcal{F}(\omega)|$ for a given flame, but has relatively small impact on $\phi_f(\omega)$. 
\begin{table}[ht]
    \centering
    \caption{Impulse response coefficients $h_k$ for DTD model}
    \begin{tabularx}{0.7\textwidth}{c| *{8}{Y}}
        \toprule
         $k$   & 2 & 3 & 4 & 5 & 6 & 7 & 8 & 9\\ \cmidrule(lr){1-9}
         $h_k$ & $0.1$ & $0.3$ & $0.6$ & $0.3$ & $0.1$&$-0.1$& $-0.2$&$-0.1$\\ 
         \bottomrule     
    \end{tabularx}
    \label{tab:DTDPar}
\end{table}

\cmnt{In an experiment where only the FFR is available, the corresponding FTF could be generated from the FFR by extrusion, Taylor expansion, or filtering \cite{SchmiBlume13, Tay-WBombe12, SubraBlume15, Polif20, BrokoFourn22}. }

\subsection{Dispersion relation of the thermoacoustic system}

The interaction between flame and acoustics in a resonator may be summarized by the corresponding dispersion relation. For the setup illustrated in Fig. \ref{fig:DuctSchematics}, the dispersion relation is given by \begin{equation}
    \xi\underbrace{\frac{\cos \varphi_u(\omega)}{\sin \varphi_u(\omega)}}_{-iZ_u(\omega)}  - (1+\theta \mathcal{F}(\omega)) \underbrace{\frac{\sin \varphi_d(\omega)}{\cos \varphi_d(\omega)}}_{-iZ_d(\omega)}  =0 \label{eq:charEq}
\end{equation}
with $\varphi_{u}(\omega)=\omega x_f/L$, $\varphi_{d}(\omega)=\omega (1-x_f/L)$, and $Z_{[u,d]}(\omega) = p^*_{[u,d ]}/u_{[u,d]}$ being the frequency dependent acoustic impedance of the flame upstream and downstream (see \cite{YongSilva21} for detailed derivations). 
On the one hand, this dispersion relation could be solved for the eigenfrequencies $\omega$ and thus the eigenmodes, if the FTF is known. 
On the other hand, it could also be solved for the FTF, if the mode eigenfrequency $\omega$ is given, 
\begin{equation}
    \mathcal{F}(\omega) = \frac{1}{\theta}\left(\xi\frac{Z_u(\omega)}{Z_d(\omega)}-1\right). \label{eq:charEq_FTF}
\end{equation}
\begin{figure}[ht]
\centering
\includegraphics[width=0.45\textwidth]{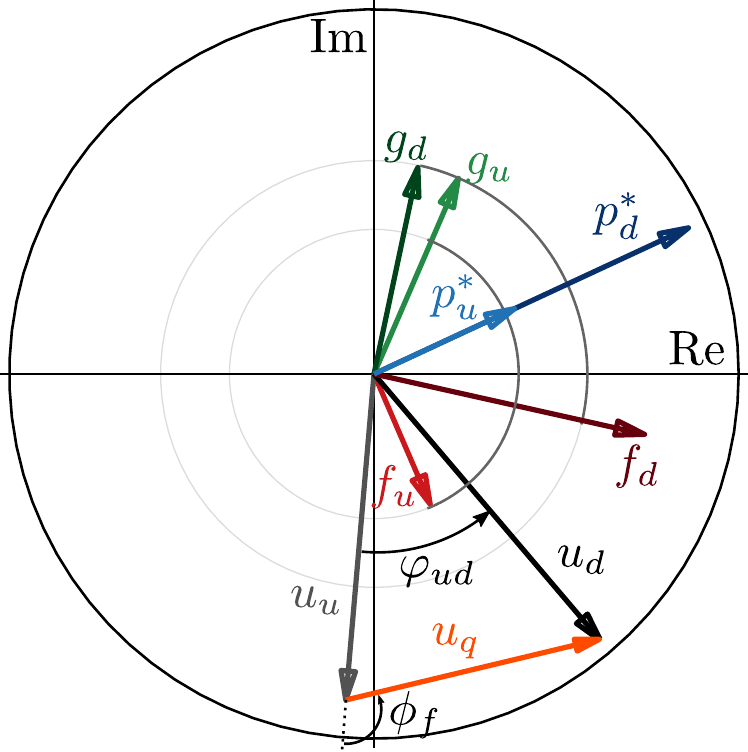}
\caption{2D phasor diagram depicting the phasors of an unstable mode ($\sigma>0$) at the upstream $[\ ]_u$ and the downstream $[\ ]_d$ of the flame at $x=x_f = 0.4L$, with the FTF gain and phase being $|\mathcal{F}| = 0.2$ and $\phi_f=0.6\pi$, respectively. The concentric circles act as a guide to highlight the difference in magnitude between the $f$ and $g$. The non-symmetrical $f,g$ phasors result in a phase difference between $u_u$ and $u_d$, $\varphi_{ud}\in (0,\pi)$. } 
\label{fig:nonMargin2D}
\end{figure}

\begin{figure}[ht]
\centering
\begin{subfigure}{.45\textwidth}
  \centering
  \includegraphics[width=\linewidth]{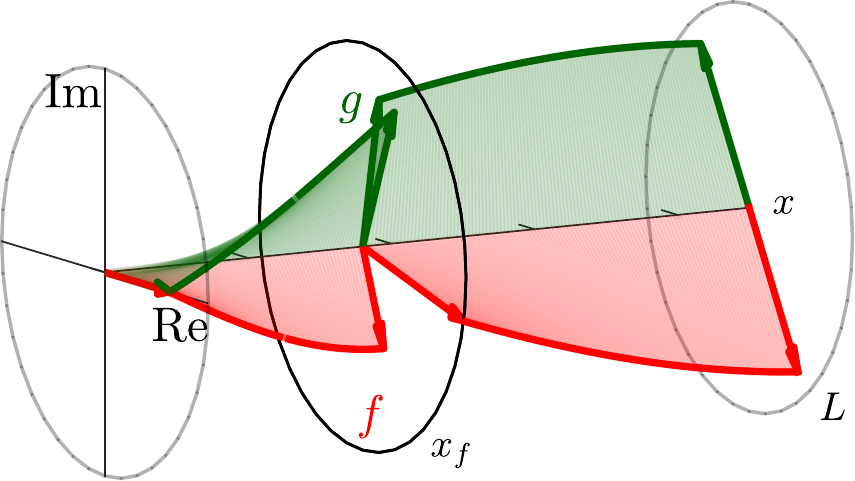}
  \caption{}
\end{subfigure}%
\hspace{5mm}
\begin{subfigure}{.45\textwidth}
  \centering
  \includegraphics[width=\linewidth]{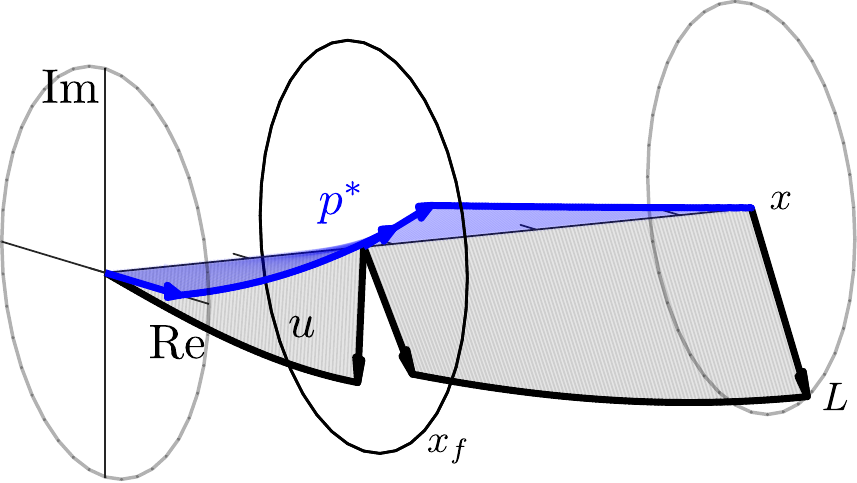}
  \caption{}
\end{subfigure}
\caption{3D phasor diagram of (a) $f,g$ and (b) $p^*,u$ for an unstable thermoacoustic mode $\sigma>0$, with FTF gain $|\mathcal{F}| = 0.13$ and phase $\phi_f=0.6\pi$, and $x_f = 0.4L$. Within the system boundaries, the $f,g$ phasors are non-symmetric, due to an amplitude decay of $f$, while simultaneously an amplitude growth of $g$ in the $+x$ direction. As a result, the $p^*,u$ phasors are not in quadrature to each other.} \label{fig:nonMargin3D}
\end{figure}

The phasor diagram in Fig. \ref{fig:nonMargin2D} depicts, by way of example, the phasors of acoustic variables immediately upstream and downstream of the flame for a growing mode $\sigma>0$ ($\textnormal{RI}>0$) \cmnt{in the ideal closed-open resonator depicted in Fig. \ref{fig:DuctSchematics}}. In contrast to marginally stable modes analyzed in \cite{YongSilva21}, the relative phase between the velocity phasors $u_u,u_d$ is not restricted to $\varphi_{ud}=0$ or $\varphi_{ud}=\pi$, but may have intermediate values $\varphi_{ud}\in[-\pi,\pi]$ determined by the growth rates.  The phasor plot also highlights the asymmetrical $f,g$ phasors (w.r.t. the real axis) and correspondingly the non-perpendicular $p,u$ phasors (compare these features against the phasors arrangement for a case $\sigma=0$ as in \cite{YongSilva21}). For illustrative purpose, the 3D phasor plot is given in Fig. \ref{fig:nonMargin3D}. 

\section{Thermoacoustic modes with infinite growth or decay rates in an ideal resonator}
In this section, we briefly recapitulate the ITA categorization of marginally stable modes as proposed in \cite{YongSilva21}. Subsequently, we investigate the phasors characteristics of extreme cases in the same ideal closed-open cavity, i.e. modes with extreme growth/decay rates $\sigma\rightarrow\pm\infty$. We will show that such modes may be categorized as ITA or acoustic modes using the existing criterion. These observations will set the basis for the general case in the next section.

\subsection{Marginally stable modes}
In our previous work \cite{YongSilva21}, the marginally stable modes in an ideal resonator, i.e. modes with real-valued frequencies $\omega\in \mathbb{R}$, were analyzed in terms of the velocity phasors. It was demonstrated that due to the symmetric rotation of the $f,g$ phasors (as discussed), the velocity phasors before and after the flame $u_u,u_d$ must be either perfectly aligned or anti-aligned. The modes with perfectly aligned $u_u,u_d$ are categorized as acoustic modes. In contrast, the modes with perfectly anti-aligned $u_u,u_d$ are categorized as ITA modes due to their resemblance to pure ITA modes, i.e. thermoacoustic modes in an anechoic environment. On the frequency axis, acoustic modes continuously transition into ITA modes and vice versa under a continuous variation of the FTF magnitude (or FFR gain, see discussions above), c.f. Fig. 14 in \cite{YongSilva21}. 

\subsection{Extremely growing modes}
At positive growth rates $\sigma>0$, the amplitudes of the outgoing acoustic waves generated at the flame $f_d,g_u$ are larger than the incoming waves $g_d,f_u$. Recall from the previous section that the resonator and the boundaries are regarded as ideal, which allows the acoustic waves generated by the flame to propagate without dissipation to and from the boundaries. Thus, the incoming waves, say, $g_d$ in the downstream cavity, may be regarded as outgoing waves $f_d$ generated at earlier times, which were reflected back at the downstream termination $x=L$, c.f. Fig. \ref{fig:fgDist}.
Thus, we may formulate
\begin{equation}
    g_d(t) = - f_d\left(t-2\tau_d\right) \quad\Rightarrow\quad  |g_d| = |f_d|\ e^{-2\sigma\tau_d} < |f_d| \label{eq:fgd_pInf}
\end{equation}
where $\tau_d = {(L-x_f )}/{c_d }$.
Correspondingly, in the upstream cavity
\begin{equation}
    f_u(t) = g_u\left(t-2\tau_u\right) \quad\Rightarrow\quad  |f_u| = |g_u|\ e^{-2\sigma\tau_u} < |g_u| \label{eq:fgu_pInf}
\end{equation}
where $\tau_u = {x_f }/{c_u }$.
In the limit of $\sigma\rightarrow+\infty$, the incoming waves $f_u,g_d$ vanish, as would be the case in an anechoic environment. Recalling Eq.~\eqref{eq:fg2pu}, the acoustic variables $p^*,u$ immediately upstream and downstream of the flame must satisfy
\begin{equation}
    \begin{split}
        p_d^*=f_d+g_d = f_d, &\quad u_d = f_d - g_d = f_d\\
        p_u^*=f_u+g_u = g_u, &\quad u_u = f_u- g_u = -g_u 
    \end{split}
\quad\Leftrightarrow\quad
    \begin{split}
        p_d^* &= u_d\\
        p_u^* &= -u_u
    \end{split}
    \label{eq:pu_pInf}
\end{equation}
Inserting Eq.~\eqref{eq:pu_pInf} into the pressure coupling relation given in Eq.~\eqref{eq:RH}, we obtain
\begin{equation}
u_d = -\xi u_u. \label{eq:u_pInf}
\end{equation}
As shown above, the infinite growth rate implies a flipping of velocity phasor across the flame. Generalizing the categorization established in our previous study, we conclude that extremely growing thermoacoustic modes should be regarded as ITA modes. 
\begin{figure}[ht]
\centering
\begin{subfigure}{.4\textwidth}
  \centering
  \includegraphics[width=\linewidth]{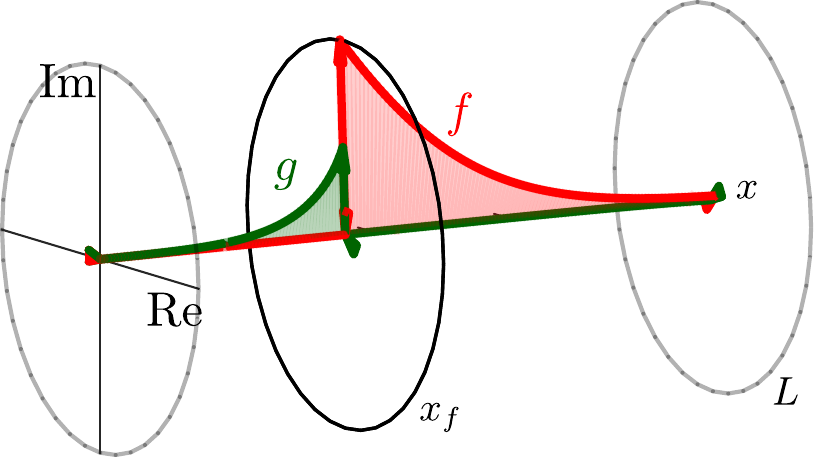}
  \caption{}
\end{subfigure}%
\hspace{5mm}
\begin{subfigure}{.4\textwidth}
  \centering
  \includegraphics[width=\linewidth]{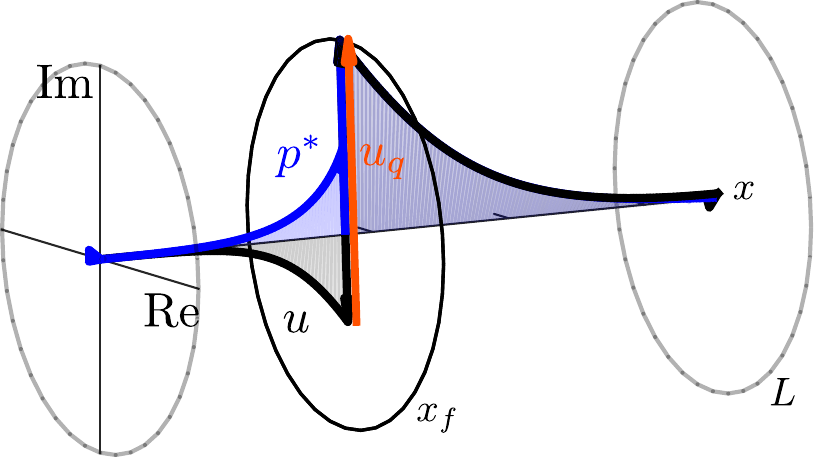}
  \caption{}
\end{subfigure}
\caption{3D Phasor diagrams of an ITA mode with $\sigma = 10^3$ and $x_f = 0.4L$, depicting the evolution of (a) $f,g$ phasors and (b) $p^*,u$ phasors in the $+x$ direction in an ideal closed-open cavity. The dominant outgoing CWAs $f_d,g_u$ and negligible incoming CWAs $f_u,g_d$ are visualized in (a). As evident in (b), the velocity phasor at the immediate flame downstream $u_d$ is almost at $\pi$-radians phase difference with respect to the upstream velocity $u_u$. At the same time, the heat release phasors $u_q$ is almost aligned with the pressure fluctuation $p^*$ to generate a maximum acoustic energy. Note that the phasor diagrams of the pure ITA mode at the same frequency are completely identical. \cmnt{The animation showing the medium oscillation for this mode is provided as supplementary material$^{\ref{fn:pureITA}}$ following the discussions in Sec. \ref{sec:displacement}}}
\label{fig:pInf3D}
\end{figure}

Substituting Eq.~\eqref{eq:u_pInf} into Eq.~\eqref{eq:RH}, we derive the FTF magnitude $|\mathcal{F}(\omega)|$ at the limit of $\sigma\rightarrow+\infty$. 
\begin{align}
\quad &\lim_{\sigma\rightarrow+\infty}\mathcal{F}(\omega)= -\frac{\xi+1}{\theta} \in \mathbb{R}^-, \, \omega\in\mathbb{C} \label{eq:FTF_infty} \\
\Rightarrow \quad&  |\mathcal{F}(\omega)| =\frac{\xi+1}{\theta}, \quad  \phi_f(\omega) = (2m+1)\pi \label{eq:magPhase}
\end{align}
\cmnt{With this FTF, we can show that} the unsteady heat release rate $\dot{Q}$, which is represented by $u_q$
\begin{align}
    u_q = u_u\theta\mathcal{F}(\omega) =  -(1+\xi) u_u
\end{align}
is perfectly in-phase with the pressure fluctuation (c.f. Fig. \ref{fig:pInf3D}), \cmnt{thus satisfying} the Rayleigh criterion for a thermoacoustic instability
\begin{align}
    \textnormal{RI} &\equiv p^*_u\cdot u_q \nonumber\\
                    &= (-u_u)\cdot (-(1+\xi) u_u) \nonumber\\
                    &= (1+\xi) |u_u|^2 > 0.
\end{align}

To achieve the high growth rates, the strength of the flame response, which is represented by the FFR gain must be infinitely large. Take the $n-\tau$ model for example, where the FFR gain is directly proportional to the growth factor $e^{\sigma\tau}$.
\begin{equation}
    |\mathcal{F}(\omega)| = ne^{-\sigma\tau} = \frac{\xi+1}{\theta} \quad \Leftrightarrow \quad n = \frac{\xi+1}{\theta}e^{\sigma\tau} \label{eq:n_scaling}
\end{equation}
Evidently, the growth of acoustic waves is not directly associated with the FTF magnitude, but with the gain of the FFR. If a flame with an extremely high gain responds in the opposite phase with respect to the upstream velocity perturbation, instability with extremely high growth rate will ensue (as expected).

\subsection{Extremely decaying modes \label{sec:XdampedITA} }

Another limiting case is that of extremely decaying modes $\sigma\rightarrow-\infty$. In this case, we find a complete annihilation of acoustic energy at the flame, such that no acoustic waves are transmitted or produced. The amplitude of the outgoing acoustic waves becomes negligibly small $f_d,g_u \rightarrow 0$. 
In analogy to the previous subsection, one derives from \eqref{eq:fgd_pInf}-\eqref{eq:fgu_pInf}
\begin{equation}
p_d^*=-u_d, \quad p_u^*=u_u; \label{eq:pu_nInf}
\end{equation}
and the coupling relation 
\begin{equation}
    u_d = -\xi u_u,
\end{equation}
which is identical to that given in Eq. \eqref{eq:u_pInf}. 
A change in sign of the velocity fluctuations across the flame is expected, thus strongly decaying modes are as well ITA modes. We obtain the FTF magnitude and phase $|\mathcal{F}(\omega)|= (\xi+1)/\theta, \, \phi_f=(2m+1)\pi$, which is identical to Eq. \eqref{eq:magPhase}. Although the FTF is identical to that in the extremely unstable case, we are not dealing with the same flame as discussed. In fact, the FFR gain approaches 0 here, which is however scaled up by the high decay rate $\sigma\rightarrow -\infty$ (see \eqref{eq:n_scaling}). Physically, this case describes an annihilation of the pressure buildup at the flame by the reduction in heat release rate fluctuation due to the diminishing flame gain.

\begin{figure}[ht]
\centering
\begin{subfigure}{.4\textwidth}
  \centering
  \includegraphics[width=\linewidth]{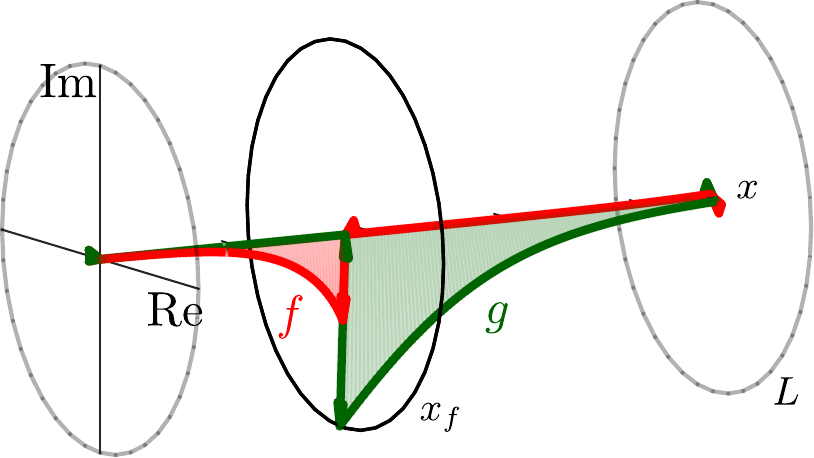}
  \caption{}
\end{subfigure}%
\hspace{5mm}
\begin{subfigure}{.4\textwidth}
  \centering
  \includegraphics[width=\linewidth]{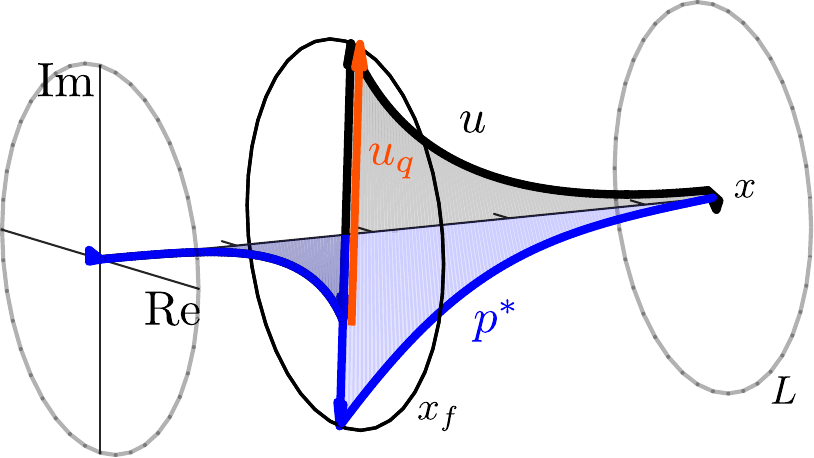}
  \caption{}
\end{subfigure}
\caption{3D Phasor diagrams of an ITA mode with large decay rate $\sigma= -10^3$ and $x_f = 0.4L$, depicting the evolution of (a) $f,g$ phasors and (b) $p^*,u$ phasors in the $+x$ direction in an ideally closed-open cavity. (a) visualizes the dominant incoming CWAs $f_u,g_d$ and negligible outgoing CWAs $f_d,g_u$. As evident in (b), the velocity phasor at the immediate flame downstream $u_d$ is almost at $\pi$-radians phase difference with respect to the upstream velocity $u_u$. At the same time, the heat release phasors $u_q$ is almost fully anti-aligned with the pressure fluctuation $p^*$ to cause maximum annihilation of acoustic energy. Note that $u_u$ is hidden by $p_u^*$ in (b).}
\label{fig:nInf3D}
\end{figure}
In Fig. \ref{fig:nInf3D}, the 3D phasor plot for an extremely decaying mode is visualized. The localized amplitudes of the incoming waves $f_u,g_d$ is immediately visible in the $f,g$ plot. Beside that, the $u_q$ phasor that is almost fully out-of-phase with $p^*$ indicates that acoustic energy is annihilated by the thermoacoustic interactions at the flame 
\begin{align}
    \textnormal{RI} &\equiv p^*_u\cdot u_q \nonumber\\
                    &= (u_u)\cdot (-(1+\xi) u_u) \nonumber\\
                    &= -(1+\xi) |u_u|^2 < 0.
\end{align}

The conclusion that extremely decaying thermoacoustic modes are generally ITA is coherent with the findings in several studies, in particular \cite{ MukheShrir17,OrchiSilva20}. It was shown that the characteristic equation in \eqref{eq:charEq} may be simplified to that of a pure ITA mode \cite{EmmerBombe15}, as the ratio of acoustic impedances $Z_u(\omega)/Z_d(\omega)$ approaches $-1$ at $\sigma\rightarrow-\infty$
\begin{equation}
    1 + \xi + \theta\mathcal{F}(\omega)=0.
\end{equation}

\section{Transition between acoustic and ITA modes with non-zero growth rates}
In the previous section, we have seen that modes in the extreme case of $\sigma \rightarrow \pm \infty$ are generally ITA. Conversely, modes on the real frequency axis consist of alternating acoustic and ITA modes \cite{YongSilva21}. This leads to the obvious questions: what happens for intermediate values of growth rates and where are transitions from acoustic to ITA modes away from the real axis? More importantly, how to define ITA modes that are not marginally stable?
The latter question is warranted, as the $u_u$ and $u_d$ phasors are not perfectly aligned or anti-aligned (see above).
In that case, such modes are not readily categorizable with the criterion outlined in our previous work \cite{YongSilva21}.

\begin{figure}[ht]
\centering
\begin{subfigure}{.5\textwidth}
  \centering
  \includegraphics[width=\linewidth]{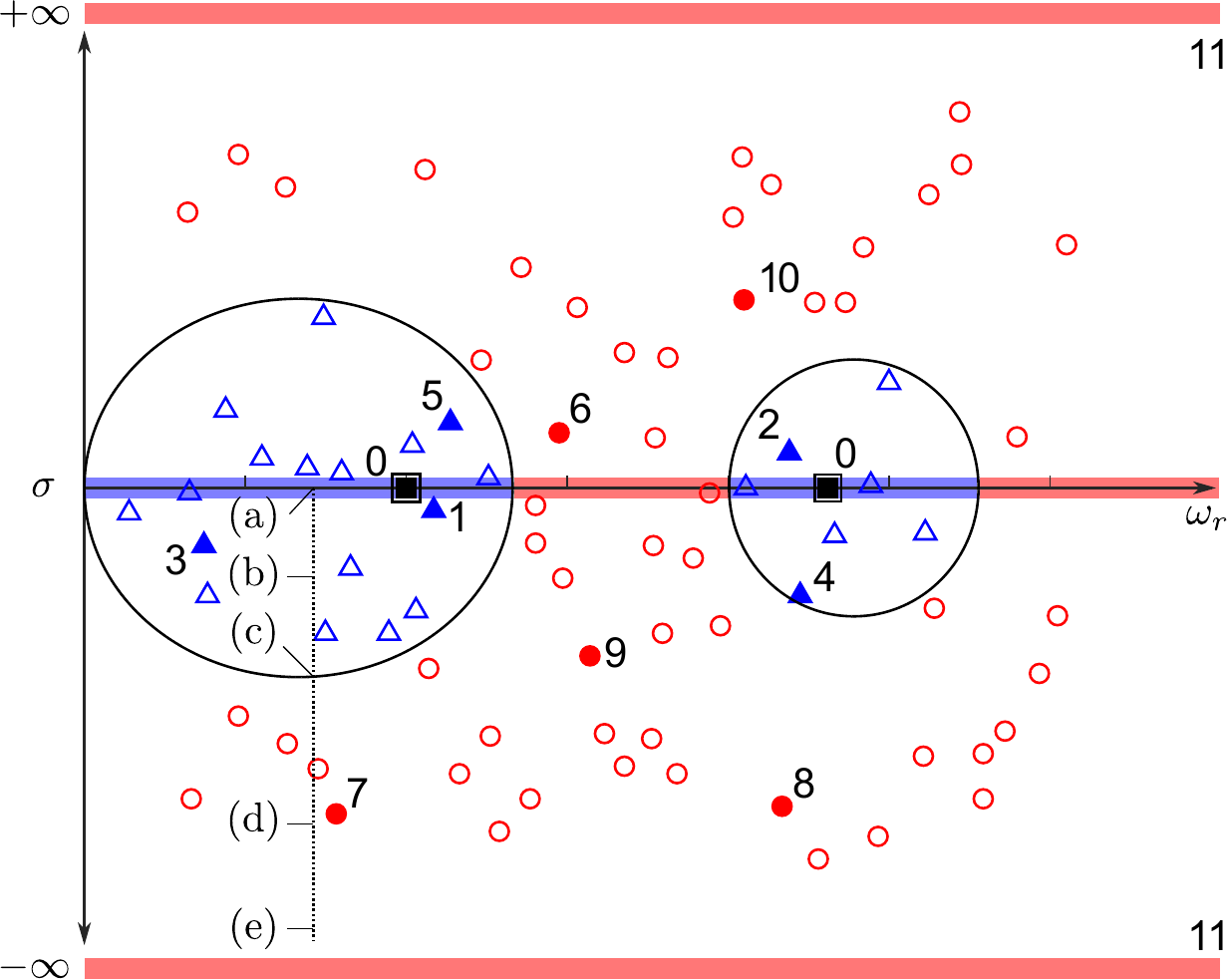}
  \caption{}
\end{subfigure}%
\hspace{10mm}
\begin{subfigure}{.4\textwidth}
  \centering
  \includegraphics[width=\linewidth]{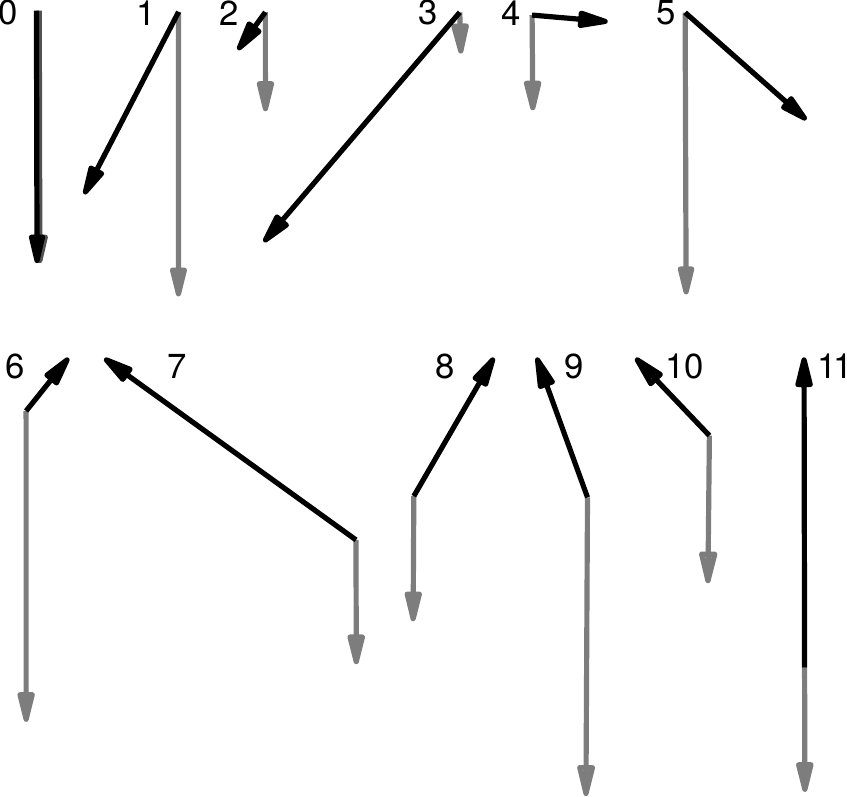}
  \caption{}
\end{subfigure}
\caption{(a) A stability map depicting the distribution of modes with partially aligned $u_u,u_d$ phasors (blue triangles) and those with partially anti-aligned $u_u,u_d$ phasors (red circles). The black squares mark the the pure acoustic modes $\omega_{p,i}$ ("0").  The red highlighted regions at $\sigma\rightarrow\pm\infty$ mark the known ITA modes in previous section, while the blue and red highlighted regions at $\sigma=0$ are known respective acoustic and ITA modes in \cite{YongSilva21}. The phasor diagrams of modes at locations labeled (a)-(e) along the vertical dotted arrow are detailed in Fig. \ref{fig:ac2ITA}. The oval patches trace the approximate areas that separate the red circles from the blue triangles. (b) Five examples of velocity phasor plots among the blue triangles ("1-5") and the red circles ("6-10") are shown. For comparison purpose, the velocity phasor plot of the pure acoustic modes ("0"), and the ITA modes at $\sigma\rightarrow\pm\infty$ ("11") are included. The grey arrow represents $u_u$ and the black arrow $u_d$. }
\label{fig:mapwVel}
\end{figure}

\begin{figure}[ht]
\centering
\begin{subfigure}{.4\textwidth}
  \centering
  \includegraphics[width=\linewidth]{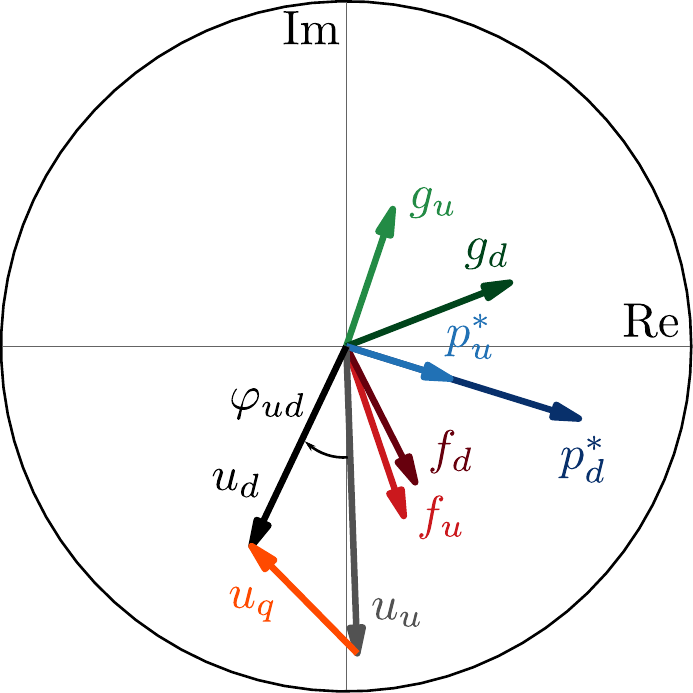}
  \caption{}
\end{subfigure}%
\hspace{5mm}
\begin{subfigure}{.4\textwidth}
  \centering
  \includegraphics[width=\linewidth]{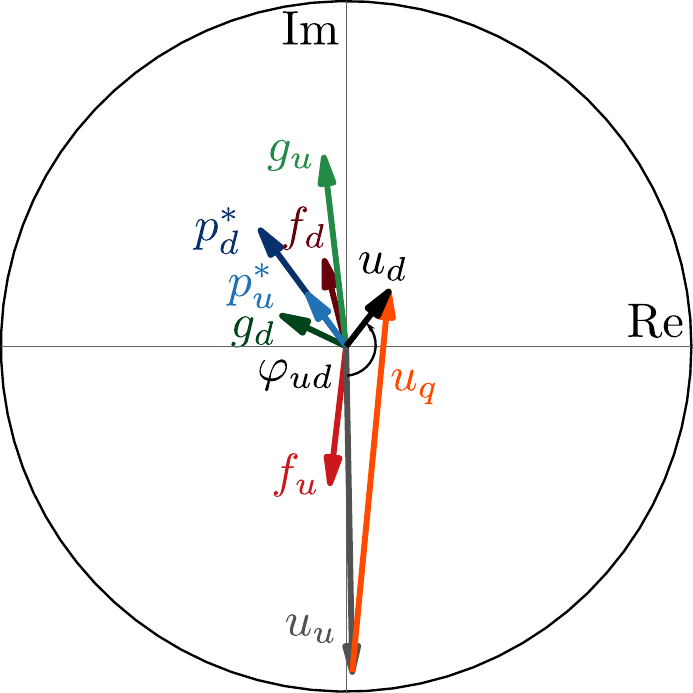}
  \caption{}
\end{subfigure}
\caption{2D Phasor diagrams depicting phasors at the immediate flame upstream $[\ ]_u$ and downstream $[\ ]_d$. (a) and (b) show the case where the velocity phasors $u_u,u_d$ are partially aligned and partially anti-aligned, respectively (c.f. Fig. \ref{fig:mapwVel}: "1" and "6"). \cmnt{The animations showing the medium oscillation for these modes are provided as supplementary materials$^{\ref{fn:AC}\ref{fn:ITA}}$ following the discussions in Sec. \ref{sec:displacement}}}
\label{fig:partTA}
\end{figure}

To identify, in a heuristic manner, a meaningful phasor characteristic for the purpose of mode categorization, a large variety of thermoacoustic modes are computed with random combinations of FTF magnitude $|\mathcal{F}(\omega)|$ and phase $\phi_f(\omega)$. 
The  dispersion relation in \eqref{eq:charEq} is solved for the corresponding complex eigenfrequencies. The relevant setup parameters are given in Tab. \ref{tab:setupPar}.
\begin{table}[ht]
    \centering
    \caption{Setup parameters for the computation of thermoacoustic modes}
    \begin{tabular}{c c}
    \toprule
         Parameter  & Value \\ \midrule
         $x_f/L$    & $0.4$\\ \addlinespace[0.01em]
         $L$          & $0.9$m\\ \addlinespace[0.01em]
         $T_u$        & $300$K\\ \addlinespace[0.01em]
         $T_d$        & $1500$K\\ \addlinespace[0.01em]
         $\theta$     & 4      \\ \addlinespace[0.01em]
         $\xi$        & $2.24$\\ \bottomrule     
    \end{tabular}
    \label{tab:setupPar}
\end{table}
With the frequencies and FTFs known, the velocity phasors $u_u,u_d$ are generated and the frequencies are plotted in a stability map, c.f. Fig. \ref{fig:mapwVel}. Note that only a handful of these phasor plots are depicted in the figure (labeled with numbers) to limit clutter. The stability map reveals:
\begin{enumerate}
\item[(i)] groups of modes in the vicinity of pure acoustic modes $\omega_{p,i}$ \cmnt{-- modes for the passive flame case --} on the neutral curve with partially aligned velocity phasors $u_u,u_d$ (represented by the blue triangles) ; 
\item[(ii)] modes outside of the groups in (i), with partial anti-alignment of $u_u,u_d$ (represented by the red circles).
\end{enumerate} 
Generalizing this observation, we categorize the former as acoustic modes and the latter as ITA modes. At the transition region between acoustic and ITA modes, $u_u$ and $u_d$ phasors are perpendicular to each other. 
For illustrative purpose, the full 2D phasor diagrams of an acoustic mode ("1" in Fig. \ref{fig:mapwVel}) and an ITA mode ("6" in Fig. \ref{fig:mapwVel}) are visualized in Fig. \ref{fig:partTA}.

\begin{figure}[ht]
\centering
\begin{subfigure}{.35\textwidth}
  \centering
  \includegraphics[width=\linewidth]{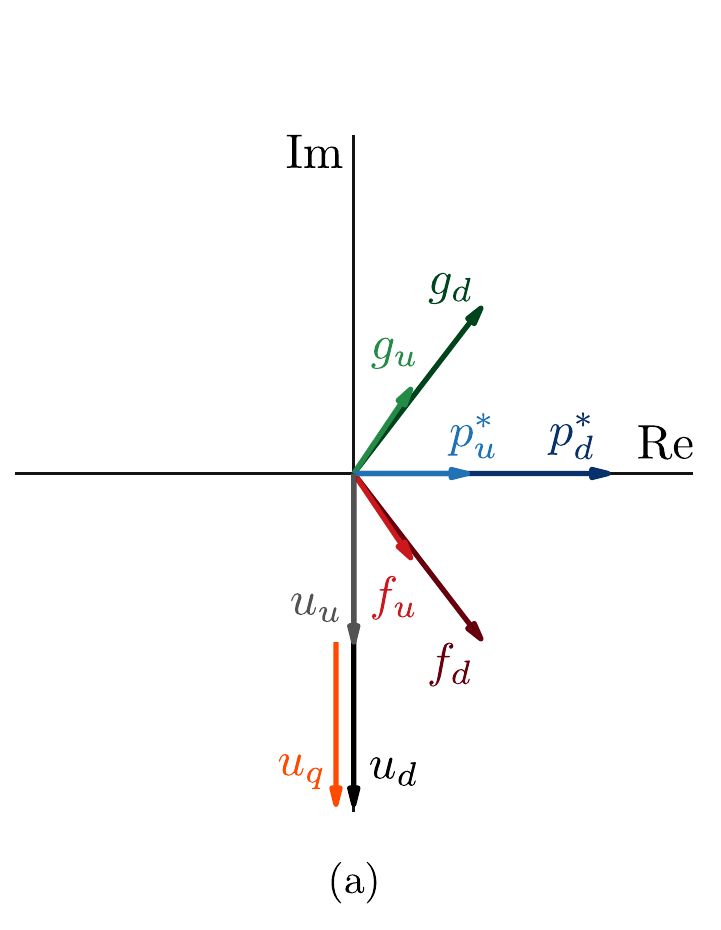}
\end{subfigure}%
\hspace{5mm}
\begin{subfigure}{.45\textwidth}
  \centering
  \includegraphics[width=\linewidth]{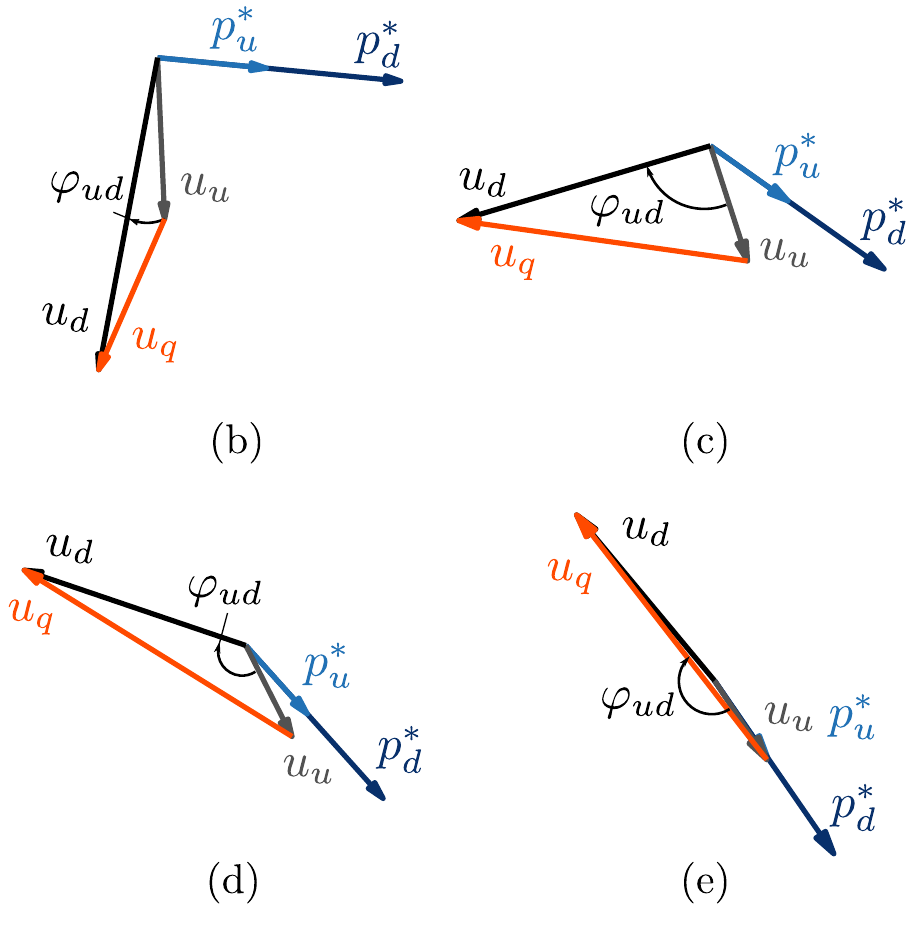}
\end{subfigure}
\caption{The phasor diagrams for thermoacoustic modes with $\omega_r/\omega_{p,1}=0.85$ and $\sigma/\omega_{p,1}\in[-2,0]$. The FTF is computed using  the dispersion relation in \eqref{eq:charEq_FTF}, which along with the complex frequency fully defines a mode. Besides (a), the phasor diagrams do not include the $f,g$ phasors to reduce cluttering. 
(a) shows the full phasor diagram of a marginally stable acoustic mode ($u_u,u_d$ in-phase, $p\perp u$). (b) shows a decaying acoustic mode ($\varphi_{ud}<\pi/2$). (c) shows a transitional mode ($\varphi_{ud}=\pi/2$). (d) shows a decaying ITA mode ($\varphi_{ud}>\pi/2$). (e) shows a highly decaying ITA mode, c.f. Sec. \ref{sec:XdampedITA}}.
\label{fig:ac2ITA}
\end{figure}

In an effort to closer investigate the mode transition from acoustic to ITA,  we take an acoustic mode on the real axis and \cmnt{judiciously modify} the FTF such that the growth rate decreases. This yields a set of thermoacoustic modes with the same oscillation frequency $\omega_r$ but different growth rates $\sigma\in(-\infty,0]$. The corresponding FTFs (magnitude and phase) are computed using Eq. \eqref{eq:charEq_FTF}. The resulting phasor diagrams are displayed in Fig. \ref{fig:ac2ITA}. The locations in a stability map of each mode in Fig. \ref{fig:ac2ITA}(a)-(e) are labeled correspondingly in Fig. \ref{fig:mapwVel} along the dotted vertical line.
As the growth rate decreases, $u_q$ phasor rotates clockwise, such that the phase difference between itself and the pressure fluctuation $p_{[u,d]}$ grows -- a direct consequence of a reduction of Rayleigh Index  (as discussed). Consequently, $u_d$ rotates in the same direction, which increases its phase deviation to $u_u$, $\varphi_{ud}>0$. Fig. \ref{fig:ac2ITA}(c) marks the transition of acoustic to ITA mode, where $u_u,u_d$ -- which are 90 degrees apart -- switch from being partially aligned to partially anti-aligned. The rotation continues allowing $u_u,u_d$ to reach full anti-alignment at high decay rates, at which the rotation stalls. This observation elucidates that the transition of the $u_u,u_d$ phasors from aligned to anti-aligned is continuous in the complex plane. 
An analogous trend may be observed, if the growth rate is increased from $0$ to $+\infty$ (not shown). In this case, instead of a clockwise rotation, $u_q$ rotates counterclockwise to increase instability, which causes a similar change observed in Fig. \ref{fig:ac2ITA}. Note that the direction of rotation of $u_q$ depends on the phase lag between the pressure and velocity phasors of the marginally acoustic mode. In this example given in Fig. \ref{fig:ac2ITA}(a), $u$ lags $p^*$. Hence, a clockwise rotating $u_q$ increases stability and vice versa. 

\section{Criterion for ITA modes with non-zero growth rates}\label{sec:criterion}

In consideration of the above findings, we define a general criterion -- $\criterionName$ criterion -- based on the scalar product between the velocity phasors $u_u$ and $u_d$ \footnotemark
\begin{equation}
\criterionName \equiv \frac{u_u}{|u_u|}\cdot\frac{u_d}{|u_d|} = \tilde{u}_u\cdot \tilde{u}_d.  \label{eq:PhiCriterion}
\end{equation}
$0<\criterionName<1$ and $-1<\criterionName<0$ indicate acoustic modes and ITA modes, respectively. \cmnt{ $\criterionName$ is a continuous (differentiable) function in the complex plane,  c.f. \ref{apx:holomorphicProof}. Thus, any continuous variation in system parameters will result in a continuous change in eigenvalues of the system, and subsequently a mode transition from acoustic to ITA or vice versa. This property of $\Phi$ is inline with the observations made by previous authors \cite{MukheShrir19,HosseKorni18a,SogarSchmi19}, where a parametric sweep may entail a mode switching, say, from acoustic to ITA and vice versa.}
\footnotetext{In the phasor representation, the complex variables $u_u$ and $u_d$ are represented as ordinary vectors in a 2D space, allowing the scalar product operation. Mathematically, the scalar product of two phasors equals the real value of the inner product between the complex variables in the complex vector space
\begin{equation}
    \criterionName=\Re{\left<\tilde{u}_u|\tilde{u}_d \right>} = \Re{(\tilde{u}_u^\dagger\tilde{u}_d)}
\end{equation}
where $\dagger$ indicates a complex conjugation.}

\begin{figure}[ht]
\centering
\includegraphics[width=\figwidth]{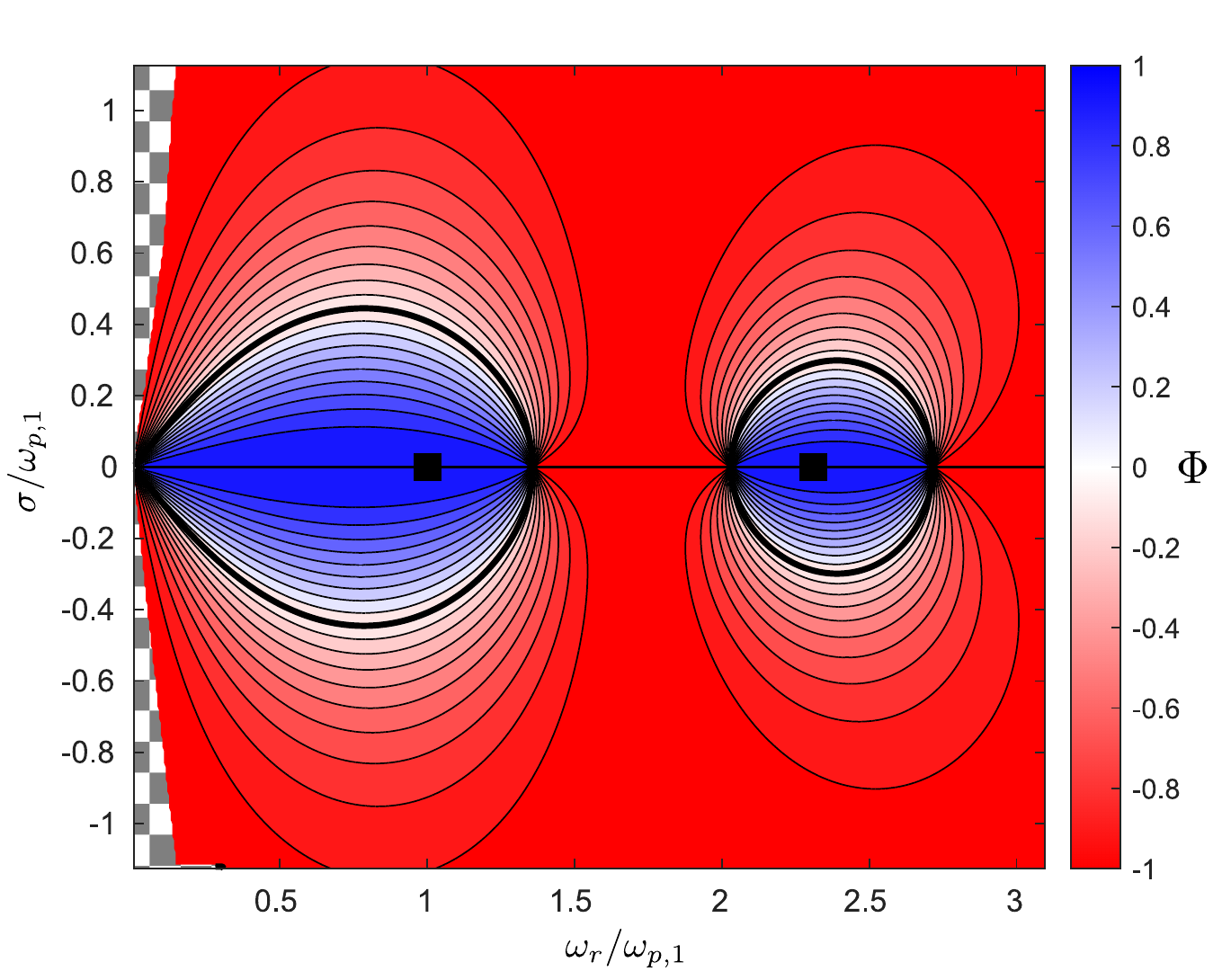}
\caption{Contour map of $\criterionName$ within the range of $\omega_r/\omega_{p,1}\in(0,3.1]$ and $\sigma/\omega_{p,1}\in[-1.2,1.2]$ for $x_f=0.4L$. The blue regions indicate acoustic modes ($\criterionName>0$) while the red region indicates ITA modes ($\criterionName<0$). The black squared markers on the neutral curve indicate the pure acoustic modes $\omega_{p,i}$. The black thickened lines tracing a semi circular path around $\omega_{p,i}$ indicate the transition of acoustic modes into ITA, and vice versa ($\criterionName=0$).  The region with checkered background is where the dispersion relation could not be easily solved due to $\omega_r\rightarrow 0$. }
\label{fig:PhiMap}
\end{figure}

Using the same computation method as that in Fig. \ref{fig:ac2ITA}, all thermoacoustic modes in the stability map within the range of $\omega_r/\omega_{p,1}\in(0,3.1]$ and $\sigma/\omega_{p,1}\in[-1.2,1.2]$ are evaluated.
The $\criterionName$ values are visualized in the contour map shown in Fig. \ref{fig:PhiMap}. The blue regions encapsulated by the oval curves ($\criterionName=0$) represent the regions of acoustic modes, while the red region contains the ITA modes. 

\begin{figure}[ht]
\centering
\includegraphics[width=\figwidth]{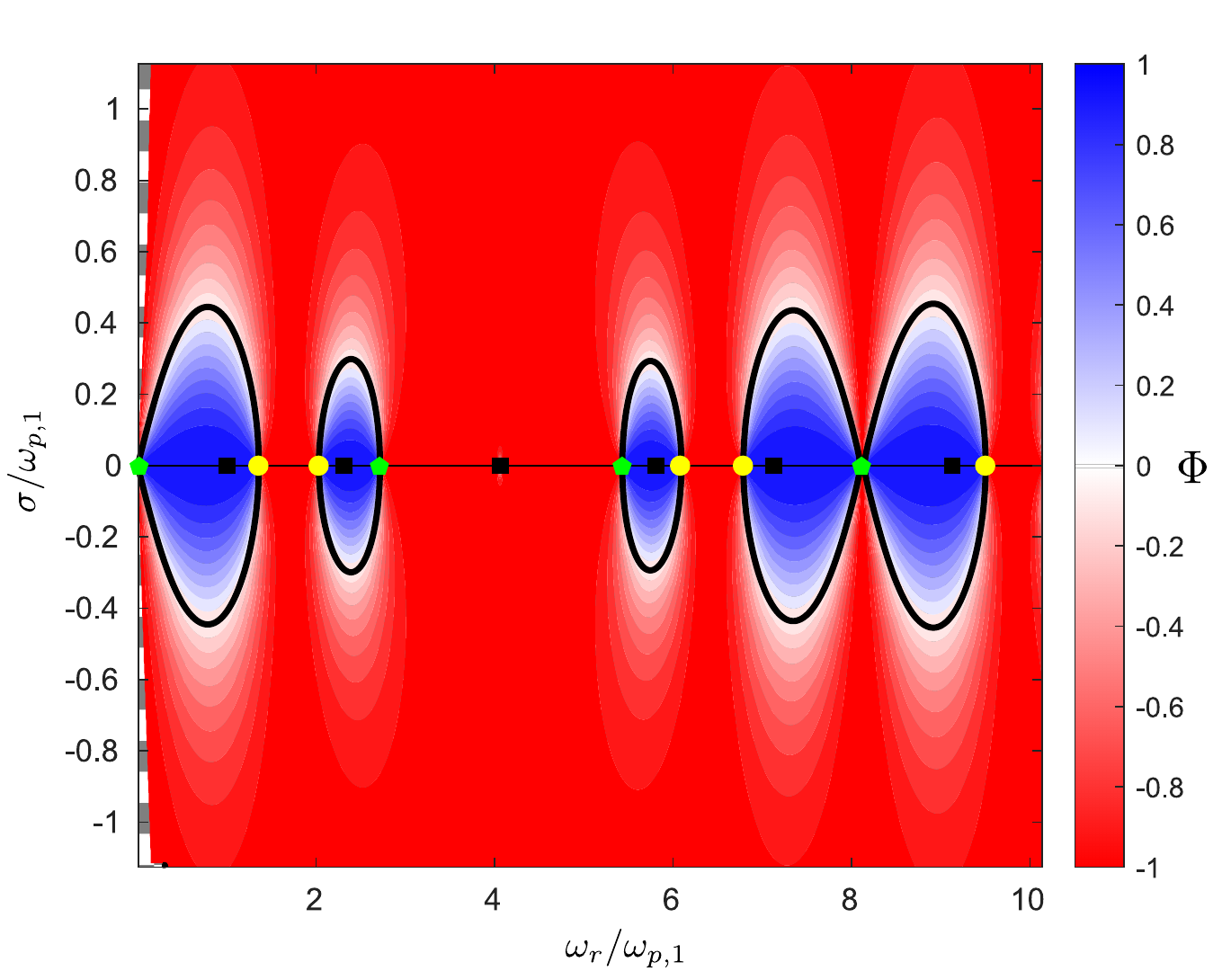}
\caption{Contour map of $\criterionName$ for larger range of $\omega_r$ compared to Fig. \ref{fig:PhiMap}. The black squares on the neutral curve indicate the pure acoustic modes $\omega_{p,i}$. The yellow circles indicate the acoustic-ITA transition of the marginally stable mode at which $\mathcal{F}(\omega) = -1/\theta$  and the green pentagons are poles of the dispersion relation given in Eq. \eqref{eq:charEq_FTF} (c.f. \cite{YongSilva21}). Other features in this map are identical to Fig. \ref{fig:PhiMap}. In this setup, ITA modes are most prevalent around $\omega_r/\omega_{p,1}=4$ (there is a small acoustic region, which is not noticeable at this scale). Note that the size and shape of the sixth acoustic region in the frequency range $\omega_r/\omega_{p,1}\approx [8.15,9.7]$ are comparable to that of the first, indicating a repetition of the previous pattern. }  
\label{fig:PhiMap_large}
\end{figure}

Note that the `islands' of acoustic modes have different sizes around the various pure acoustic modes $\omega_{p,i}$. To get an insight on how the size of the islands changes as the frequency grows, the contour map is extended to include higher values of $\omega_r$, c.f. Fig.~\ref{fig:PhiMap_large}.
The size of the acoustic island decreases initially but grows again. In this setup, we observe that the shrinking and growing of the acoustic islands repeats after $\omega_r/\omega_{p,1}\approx 8.15$, which creates an alternating pattern in the complex plane. 
ITA modes are most prevalent at around $\omega_r/\omega_{p,1}\approx 4$ (there is a very small acoustic region, which is not noticeable at this scale). 
The size of each island is determined by the distance between the acoustic-ITA transitional point of marginally stable modes (yellow dots) and the pole of the dispersion relation (green pentagon) given in Eq. \eqref{eq:charEq_FTF}, i.e when $\mathcal{F}(\omega)\rightarrow \infty$  (refer to \cite{YongSilva21} for detailed discussions). The FTF at the transitional point on the real axis is given by $\mathcal{F} = -1/\theta$ and the dispersion relation may be simplified to
\begin{align}
\cos(\omega_r\tau_u)\cos(\omega_r\tau_d) = 0
\end{align}
which yields the transition frequencies
\begin{align}
\omega_{r,\textnormal{tran}} = \frac{2m+1}{2}\frac{\pi}{\tau_{u/d}}.
\end{align}
At the poles, the dispersion relation is given by
\begin{align}
\sin(\omega_r\tau_u)\sin(\omega_r\tau_d) = 0
\end{align}
which yields the pole frequencies
\begin{align}
\omega_{r,\textnormal{pol}} = \frac{m\pi}{\tau_{u/d}}.
\end{align}
The prevalence of acoustic modes in the vicinity of the real axis, which is determined by
\begin{equation}
    \Delta \omega_r = |\omega_{r,\textnormal{pol}}-\omega_{r,\textnormal{tran}}|,
\end{equation}
is thus a function of $\tau_u$ and $\tau_{d}$, or rather the flame position $x_f$. This conclusion sheds a new light into the observation in \cite{FournSchae22}, where it was shown that the flame position inside a simple Rijke tube or a can-annular combustor could strongly influence the nature of the observed eigenmodes, and modes could easily switch nature between acoustic and ITA as the flame position varies. 

\begin{figure}[htbp]
\centering
\includegraphics[width=\figwidth]{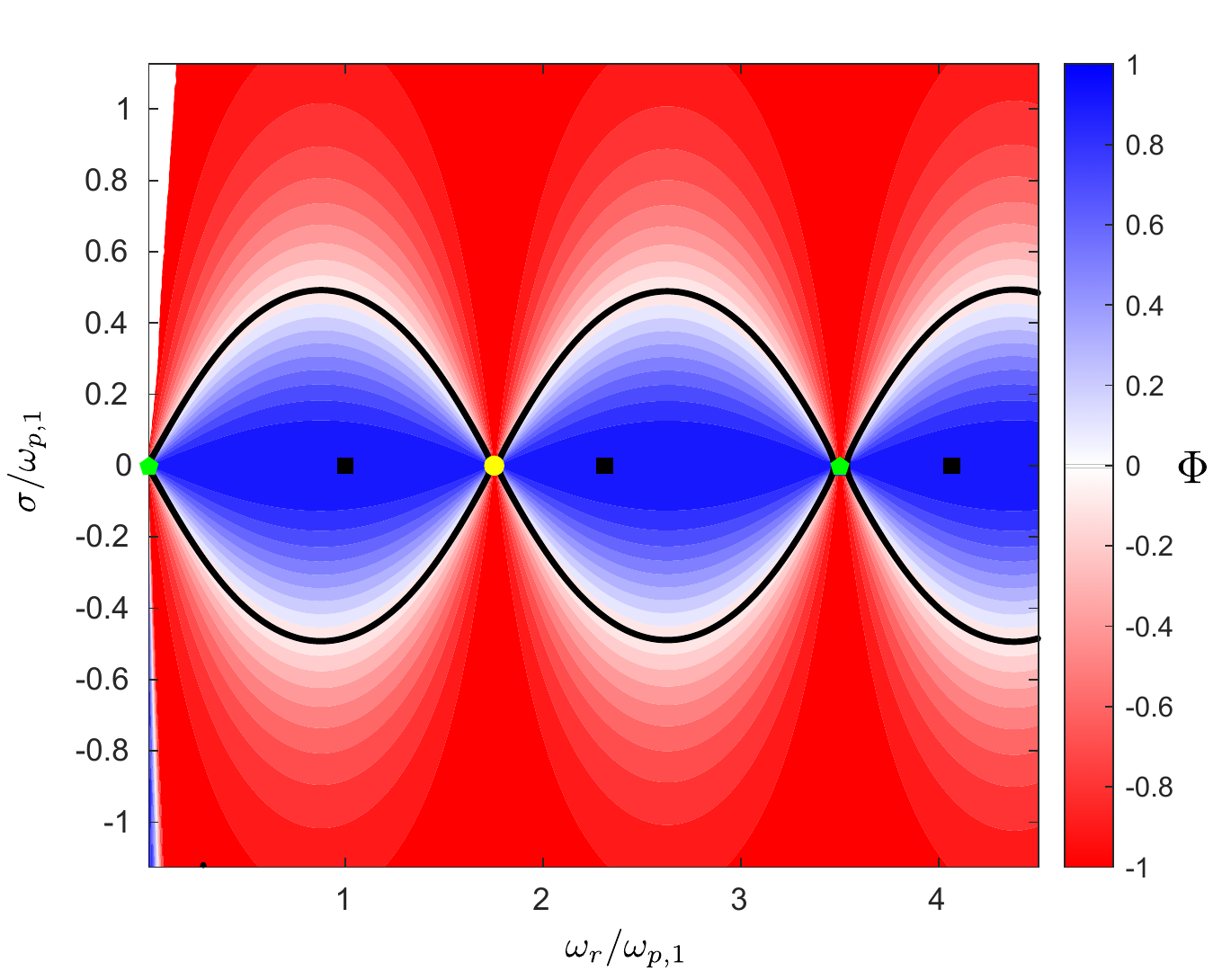}
\caption{Contour map of $\criterionName$ for $x_f=0.313L$, i.e. $\tau_u=\tau_d$. Other features in this map is identical to Fig. \ref{fig:PhiMap}.. In this case, ITA mode does not exist on the real axis. }
\label{fig:PhiMap_equitau}
\end{figure}

The acoustic regions are at maximum size with $\tau_u=\tau_d$, \cmnt{resulting in every $\omega_{r,\text{tran}}$ being exactly the average of two consecutive $\omega_{r,\text{pol}}$}. In this case, an ITA mode does not exist on the real axis, c.f. Fig. \ref{fig:PhiMap_equitau}. 
Of course, the derivations that lead to this conclusion revolves around the dispersion relation given in Eq. \eqref{eq:charEq}, which is unique to an ideally closed-open resonator. More general conclusions on resonators with different boundary conditions are not developed in this study. 

\subsection{Transitional FTF magnitude and phase}\label{sec:FTF_trans}

At the transition, the velocity phasors $u_u,u_d$ are perpendicular to each other (with the exception of the marginally stable transitional mode, where $u_d=0$)
, which results in $\criterionName=0$. Substituting the Eq.~\eqref{eq:RH} into Eq.~\eqref{eq:PhiCriterion}, yields 
\begin{equation}
u_u\cdot u_d = |u_u|^2\left(1+\theta|\mathcal{F}(\omega)|\cos(\phi_f)\right) = 0 \label{eq:FTF_criterion}
\end{equation}
The equation is satisfied only for $\cos(\phi_f)<0$, which yields the necessary transitional FTF magnitude $|\mathcal{F}_T|$ and phase $\phi_f$
\begin{equation}
|\mathcal{F}_T(\omega)|=-\frac{1}{\theta\cos(\phi_f)}, \quad \text{with } \left|\phi_f\right|\in\left(\frac{\pi}{2},\frac{3\pi}{2}\right). \label{eq:FTF_transition}
\end{equation}

\begin{figure}[ht]
\centering
\includegraphics[width=0.5\textwidth]{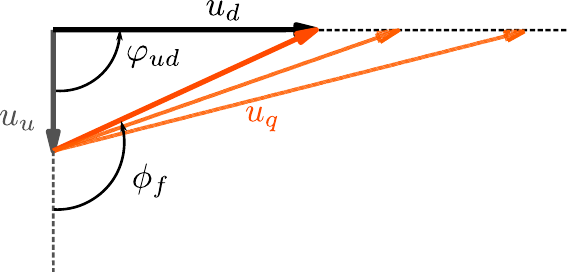}
\caption{Orientation of the velocity phasors across the flame of a mode at the acoustic-ITA transition. $u_q$ must be out-of-phase with $u_u$, while the $|\mathcal{F}(\omega)|$ must so large, so that $u_d\perp u_u$. 
}
\label{fig:uqITA}
\end{figure}

In order for an acoustic mode to transition to ITA, the volumetric oscillation caused by the active flame must be strong enough to act against the upstream displacement rate, with the flame being out-of-phase with the upstream velocity. With an FTF magnitude larger than the transitional value
\begin{equation}
    |\mathcal{F}(\omega)|>|\mathcal{F}_T(\omega)|
\end{equation}
we have an ITA mode at hand. 
Figure~\ref{fig:uqITA} shows schematically various combination of FTF gain and phase at the transition.  In the limit case where $\phi_f\rightarrow{\pi}/{2}$ or $\phi_f\rightarrow{3\pi}/{2}$, the $\mathcal{F}_T$ must be infinitely large (at the poles).  A mode with $\phi_f<\pi/2$ or $\phi_f>3\pi/2$ is always an acoustic mode. 

\begin{figure}[ht]
\centering
\begin{subfigure}{.5\textwidth}
  \centering
  \includegraphics[width=\linewidth]{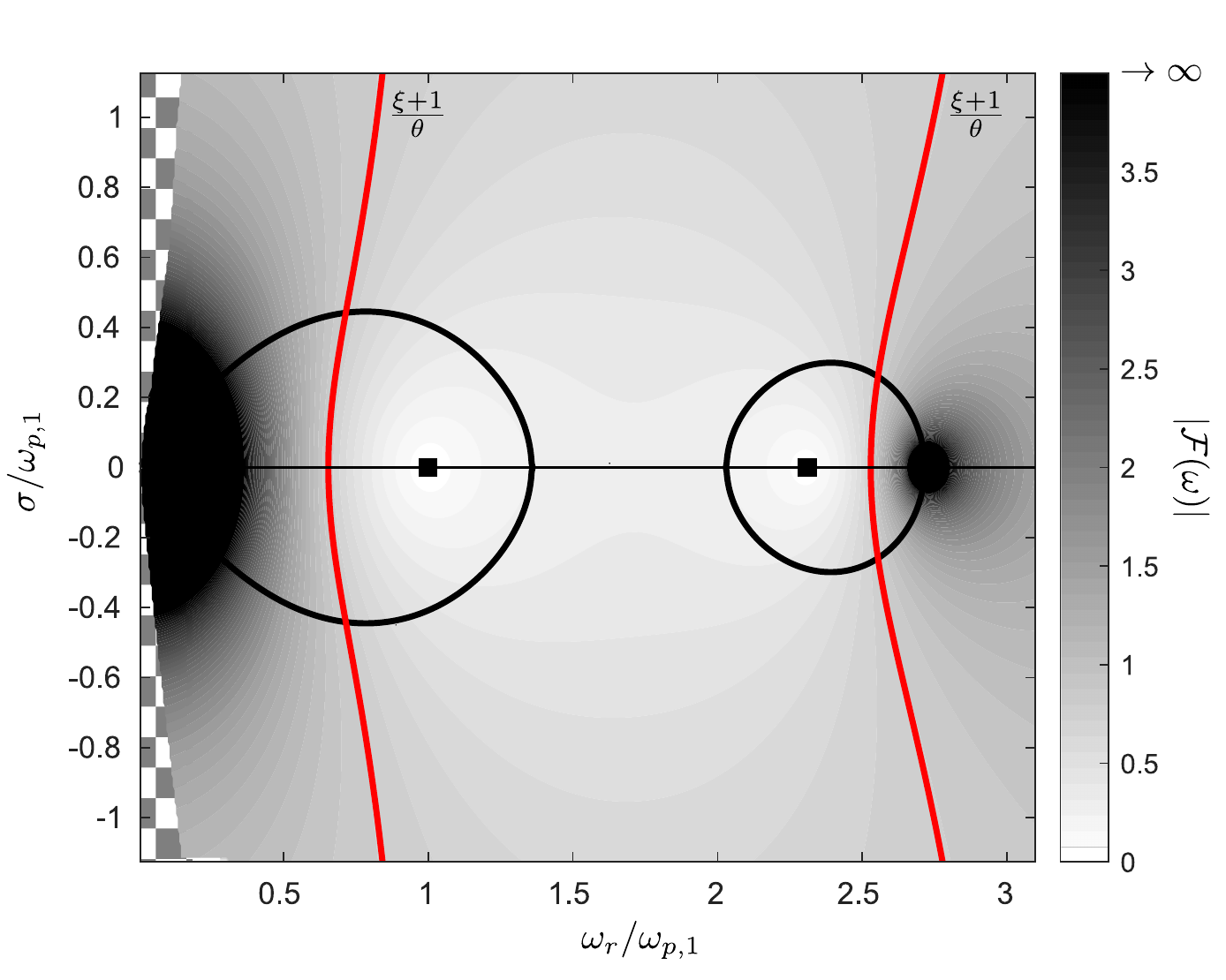}
  \caption{$|\mathcal{F}(\omega)|$}
  \label{fig:absFTFMap}
\end{subfigure}%
\begin{subfigure}{.5\textwidth}
  \centering
  \includegraphics[width=\linewidth]{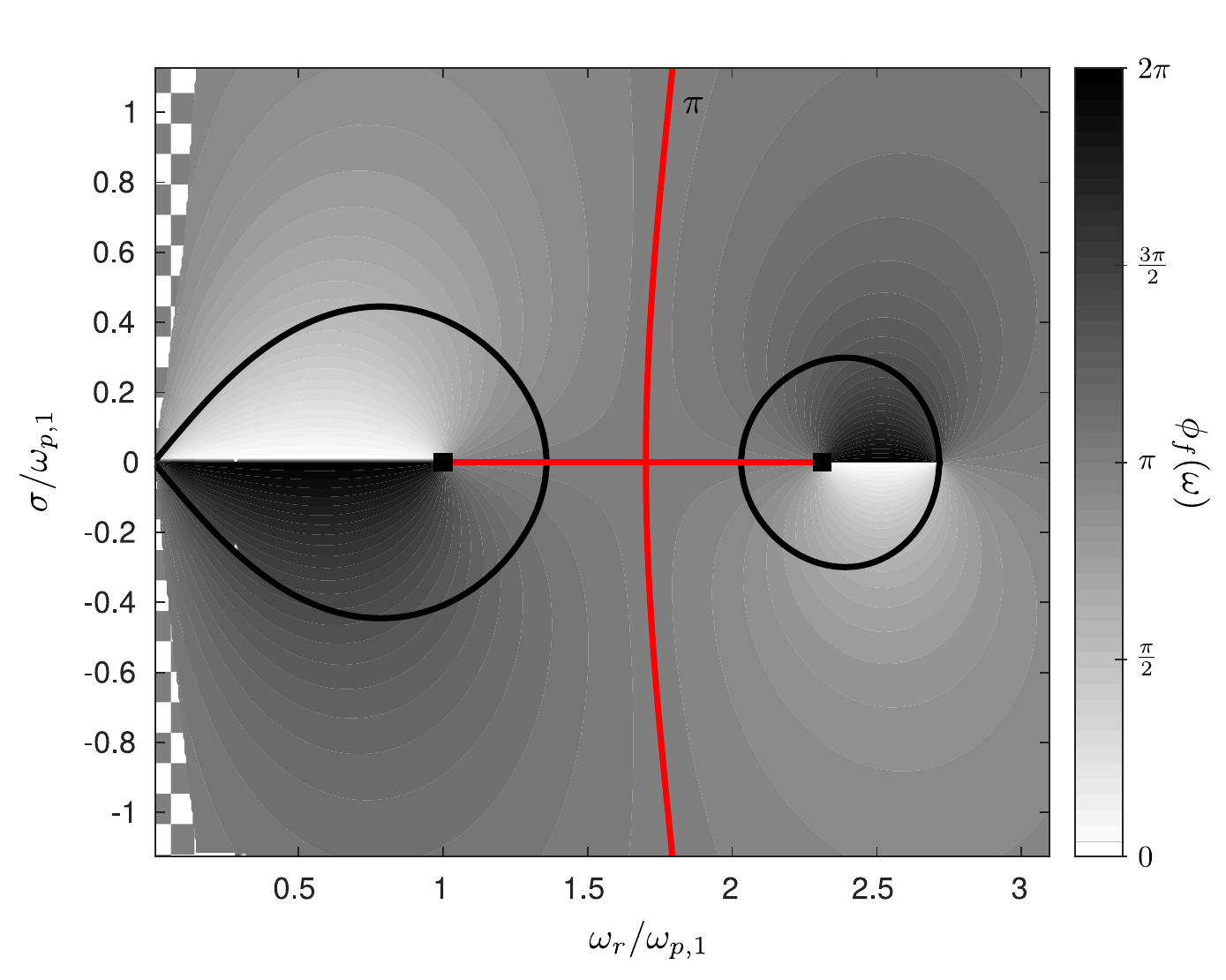}
  \caption{$\phi_f(\omega)$}
  \label{fig:phifMap}
\end{subfigure}
\caption{Contour maps of $\mathcal{F}(\omega)$ for $x_f=0.4L$.  The red line in (a) marks the value of $|\mathcal{F}(\omega)|$ that extends to $\sigma\rightarrow\pm\infty$, i.e. $|\mathcal{F}(\omega)|=\frac{\xi+1}{\theta}$, while in (b) marks the $\phi_f(\omega)$ value for marginally stable modes that also extends to $\pm\infty$, i.e. $\phi_f(\omega)=(2m+1)\pi$. The black thickened lines indicating the transition of acoustic modes into ITA and the black squared markers indicating the pure acoustic modes $\omega_{p,i}$ are included as references to Fig. \ref{fig:PhiMap}. The region with checkered background is where the dispersion relation could not be easily solved due to $\omega_r\rightarrow 0$. }
\label{fig:FTFMap}
\end{figure}

Note that the transitional FTF magnitude and phase given in Eq.~\eqref{eq:FTF_transition} are generally applicable, i.e. independent of the FTF models. For instance, we computed the FTFs (magnitude and phase) from Eq. \eqref{eq:charEq_FTF} for the modes discussed in Fig. \ref{fig:PhiMap}, without considering a specific FTF model, as shown in Fig. \ref{fig:FTFMap}. Using the FTF phase $\phi_f(\omega)$ in Fig. \ref{fig:FTFMap}(b), one may compute the transitional values $|\mathcal{F}_T(\omega)|$ and compare them against the $|\mathcal{F}(\omega)|$ in Fig. \ref{fig:FTFMap}(a) to define the acoustic and ITA regions. Naturally, the results match those identified previously. 

In practice, where a flame is described by a specific FTF model, the eigenmodes are only sparsely populated on the complex plane and therefore are a subset of the solutions presented here. As an example, the mode categorization in a test case with a flame described by a DTD model is demonstrated in \ref{apx:testCase}

\subsection{Velocity phasors as vectors of medium displacement rates}\label{sec:displacement}

The change in sign of the pressure gradient was featured in \cite{YongSilva21} as a readily discernible characteristic of the proposed categorization method. \cmnt{In this work, we will instead use the velocities of the medium (gas) to develop an intuitive understanding to the question `what does it mean for a mode to be acoustic or ITA?'. }

The real part of a velocity phasor represents the displacement rate of the medium
\begin{equation}
\dot{x}_{j}(x,t)=\Re{\left(u_{j}(x,t)\right)} = |u_{j}|e^{\sigma t} \cos(\omega_r t+\alpha_{j}), \quad j = u,d
\end{equation}
where $\alpha_{j}=\arg(u_{j})$ is the relative phase between $u_j$ phasor and the real axis. $\dot{x}_j >0$ signifies a displacement in the $+x$ direction. 

Without an active flame, the medium between two velocity nodes in a standing wave oscillates back and forth –- \emph{swings} -- at a given resonant frequency. This is coherent with the aligned $u_u,u_d$ phasors observed in pure acoustic modes\footnote{The special case, in which the flame, i.e. the region of interest, is located exactly at the velocity node is not analyzed, as they are not meaningful for the discussions in this work. 
For pedagogical purposes, it is discussed in detail in \cite{YongSilva21} in the Appendix chapters.}.
In contrast, an active flame acts as a monopole source of volume. Thus, in an anechoic environment, $u_u,u_d$ phasors are anti-aligned, which describes an inwards-outwards oscillation -- \emph{pulsation} -- at the flame. However, as shown, $u_u,u_d$ are in general partially aligned or anti-aligned, i.e. the medium at the vicinity of the flame oscillates in a superposed swinging and pulsating motion. For an acoustic mode, the swinging oscillation dominates, and vice versa for an ITA mode. 

\cmnt{Exemplary animations visualizing the oscillation of a pure acoustic\footnote{\label{fn:pureAC} PureAcoustic.mp4 } and a pure ITA mode\footnote{\label{fn:pureITA} PureITA\_Fig8.mp4 } (corresp. Fig. \ref{fig:pInf3D}), as well as an acoustic\footnote{\label{fn:AC} Acoustic\_Fig11(a).mp4 } (corresp. Fig. \ref{fig:partTA}(a)) and an ITA mode\footnote{\label{fn:ITA} ITA\_Fig11(b).mp4 } (corresp. Fig. \ref{fig:partTA}(b)) are uploaded as supplementary materials}. The mathematical description of the medium oscillations, which could lead to the derivation of the $\criterionName$ criterion \eqref{eq:PhiCriterion} is detailed in \ref{apx:oscillation}.

\section{Summary, Conclusion and Outlook}

In extension to the work of Yong et al. \cite{YongSilva21}, this study used phasor diagrams to visualize the propagation of characteristic wave amplitudes of \emph{growing} as well as \emph{decaying} thermoacoustic modes in an ideal closed-open resonator, which contains a compact, velocity sensitive flame. The inspection of the phasor diagrams suggests that the velocity phasor may shift by an intermediate phase between $-\pi$ and $\pi$ across the flame, as a result of the non-zero growth rate. 
At extremely high growth and decay rates, the velocity phasor changes direction across the flame, implying ITA modes according to the established criterion in \cite{YongSilva21}. The phasor diagrams of a variety of randomly selected modes were constructed and their corresponding locations in a stability map were determined. A clear pattern emerged:  groups of modes around the pure acoustic modes have partially aligned flame upstream and downstream velocity phasors, while modes beyond these groups have partially anti-aligned velocity phasors. The first group of modes was subsequently categorized as acoustic modes, while the latter as ITA modes. It was demonstrated that the mode transition from acoustic to ITA and vice versa is a continuous process.

Representing the acoustic variables as phasors in the ordinary vector space allow the formulation of a generalized categorization criterion -- $\criterionName$ criterion --  based on the dot product between the upstream and downstream velocity phasors, given in Eq. \eqref{eq:PhiCriterion}. A contour map depicting the distribution of acoustic and ITA modes was generated. Interestingly, the size of the acoustic region varies as the frequency increases, signifying that the prevalence of ITA modes, and correspondingly acoustic modes, in the proximity of the real frequency axis may vary for different combustor setups. Indeed, the size of the acoustic region is a function of flame position, as well as of reflective boundary conditions. \cmnt{As an extension to the $\criterionName$ criterion, the transitional FTF magnitude was derived to enable a more practical mode categorization, where the corresponding FTF is compared against the transitional value.}

Physically, the real value of a velocity phasor represents the displacement rate of the medium in a resonator. The perfectly anti-aligned velocity phasors across the flame, as typically seen in a pure ITA mode, indicate an unsteady volumetric oscillation -- `pulsating' -- in the reactive region of the flame. A pair of perfectly aligned velocity phasors, on the other hand, represents the ordinary back and forth motion -- `swinging' -- observed in a standing wave. In an echoic resonator with an active flame, the acoustic feedback is coupled with the ITA feedback mechanism. Thus, a superposition of both oscillation motions is expected. 
\cmnt{Note that the swinging and pulsating characteristics are equivalent to the transitional FTF magnitude, both of which can be derived from the $\criterionName$ criterion. Thus, a mode may be categorized by analyzing either the FTF data or the velocity measurements on the flame upstream and downstream, depending on the circumstances and available diagnostic tools in an experiment.}

In this work, we demonstrated that growing as well as decaying ITA modes could be reasonably categorized by means of phasors analysis. In addition to the known benefits of the phasor based categorization, such as straightforward implementation, physical relevance and low computational cost, the categorization of modes with partially aligned and anti-aligned velocity phasors across the flame widens its application.

In future works, more complex systems with non-ideal boundaries, internal losses or area changes, as well as can-annular geometries, shall be analyzed. In essence, we expect the $\criterionName$ criterion to be valid in these cases. Besides, the unique characteristics of ITA modes categorized this way shall be explored. It is interesting to investigate, if the peculiar characteristics observed in \cite{SilvaEmmer15,SilvaMerk17,MukheShrir17,XuZheng20} could be explained with the phasor characteristics of ITA modes. 

\section*{Declaration of Competing Interest}
The authors declared that there is no conflict of interest.

\section*{Acknowledgements}
Financial support has been provided by the German Research Foundation (Deutsche Forschungsgemeinschaft - DFG) for the SelfiXs project, DFG PO 710/21-1, in the framework of the International Collaborative Research Project Initiative ANR-DFG NLE 2018. Guillaume J. J. Fournier acknowledges the funding from the European Union's Horizon 2020 research and innovation programme under Grant Agreement No 765998 \textit{Annular Instabilities and Transient Phenomena in Gas Turbine Combustors} (ANNULIGhT).

\bibliography{YongSilva22}

%% file: Chapters/YongSilva22_Appendix.tex
\section{Test case with a DTD flame model\label{apx:testCase}}

In this section, we will demonstrate the implementation of the proposed categorization criterion in a test case with an FTF modeled by the DTD model given in Eq.~\eqref{eq:FTF_DTD}. The corresponding impulse response coefficients $h_k$ are tabulated in Tab. \ref{tab:hk_DTD}. 
\begin{table}[ht]
    \centering
    \caption{The impulse response coefficients $h_k$ of the DTD model at the corresponding element $k$. The discrete time step in this model is $\Delta t=2\times 10^{-4}\textnormal{s}$.}
    \begin{tabularx}{0.7\textwidth}{c| *{8}{Y}}
        \toprule
         $k$   & 2 & 3 & 4 & 5 & 6 & 7 & 8 & 9\\ \cmidrule(lr){1-9}
         $h_k$ & $0.1$ & $0.3$ & $0.8$ & $0.3$ & $0.1$&$-0.2$& $-0.4$&$-0.2$\\ 
         \bottomrule     
    \end{tabularx}
    \label{tab:hk_DTD}
\end{table}

\begin{figure}[ht]
\centering
\begin{subfigure}{.45\textwidth}
  \centering
  \includegraphics[width=\linewidth]{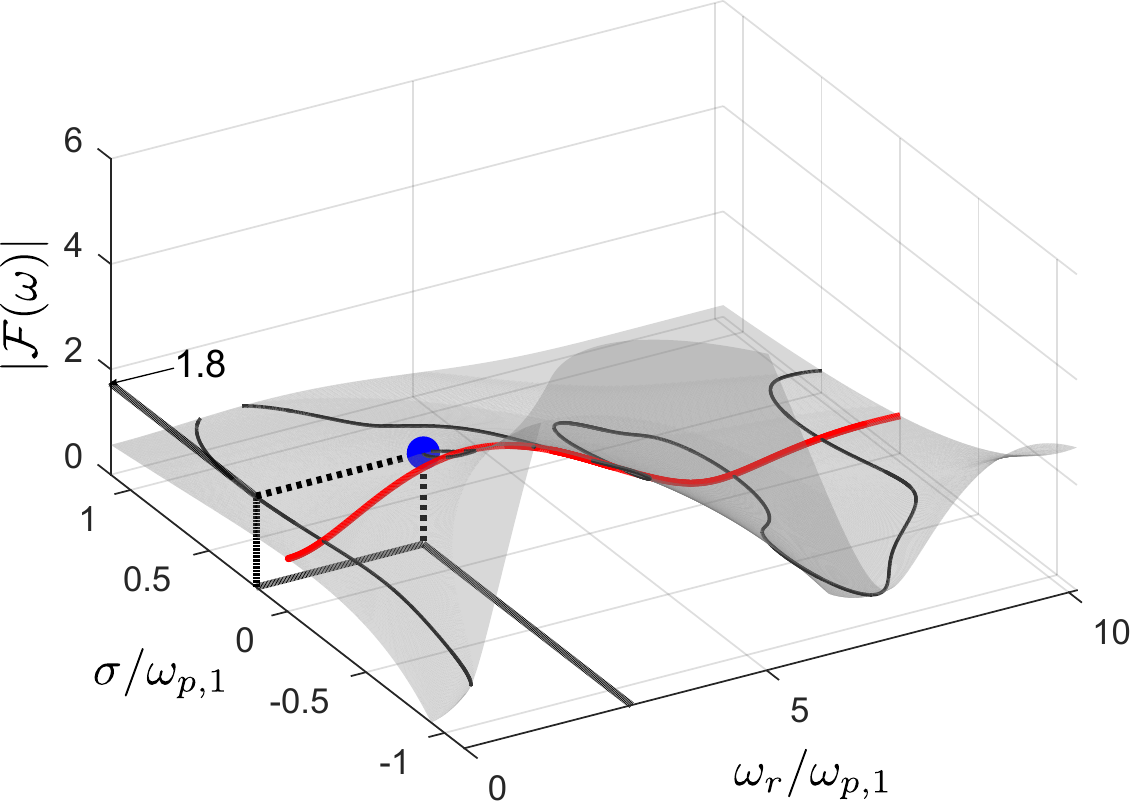}
  \caption{}
  \label{fig:absFTF_DTD}
\end{subfigure}%
\hspace{5mm}
\begin{subfigure}{.45\textwidth}
  \centering
  \includegraphics[width=\linewidth]{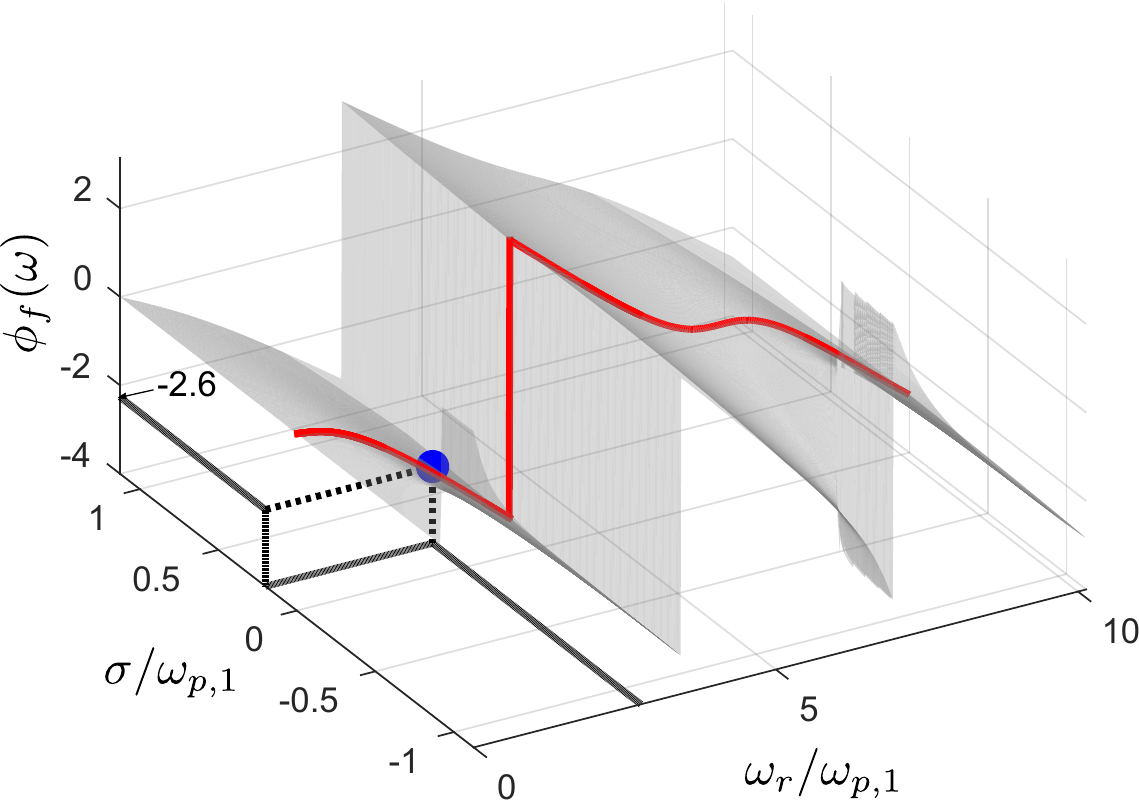}
  \caption{}
  \label{fig:phif_DTD}
\end{subfigure}
\caption{The surface plots of (a) FTF magnitude and (b) phase of a DTD flame model with excess gain and exhibit a low pass behaviour. The red curve depicts the FFR. The black curves on the surface in (a) mark the solutions that satisfy Eq. \eqref{eq:FTF_solution}. The solutions for the phase are not shown in (b) to reduce clutter. The blue filled circle is the mode of interest with $\omega_r/\omega_{p,1}=2.8$ and $\sigma/\omega_{p,1}=0.2$.  }
\label{fig:FTF_DTD}
\end{figure}

\begin{figure}[htbp]
\centering
  \includegraphics[width=0.5\textwidth]{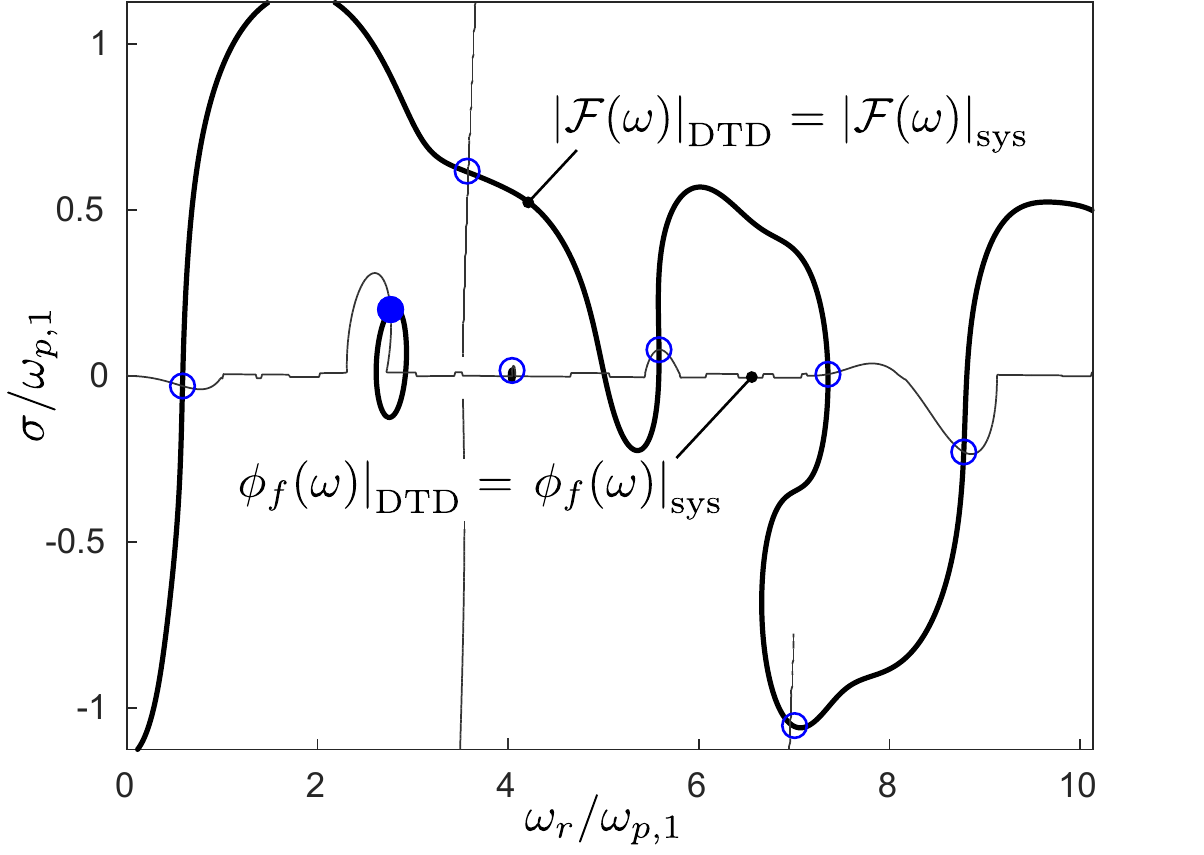}
\caption{The thickened and thin curves satisfy Eq. \eqref{eq:FTF_solution} and \eqref{eq:phif_solution} respectively. The intersection between them are all the eigenvalues to the thermoacoustic system (blue circles). The filled circle indicates the mode of interest with $\omega_r/\omega_{p,1}=2.8$ and $\sigma/\omega_{p,1}=0.2$. }
\label{fig:charSol}
\end{figure}

The resultant FTF of the DTD model in terms of magnitude and phase are visualized in Fig. \ref{fig:FTF_DTD}. This FTF must simultaneously satisfy the dispersion relation given in Eq.~\eqref{eq:charEq_FTF}, such that 
\begin{align}
    |\mathcal{F}(\omega)|_\textnormal{DTD} = |\mathcal{F}(\omega)|_\textnormal{sys} \label{eq:FTF_solution} \\ \left.\phi_f(\omega)\right|_\textnormal{DTD} = \left.\phi_f(\omega)\right|_\textnormal{sys} \label{eq:phif_solution} 
\end{align}
where the index $[\ ]_\textnormal{DTD}$ indicates the FTF of the flame model, and $[\ ]_\textnormal{sys}$ indicates the FTF computed with the dispersion relation \eqref{eq:charEq_FTF}, the results of which are visualized in Fig. \ref{fig:FTFMap}. The solutions that satisfy Eq. \eqref{eq:FTF_solution} are indicated by the black line in Fig. \ref{fig:FTF_DTD}(a). The solutions for Eq. \eqref{eq:phif_solution} are not shown to reduce cluttering.
The solutions satisfying both the Eqs. \eqref{eq:FTF_solution}-\eqref{eq:phif_solution} are the eigenvalues of the thermoacoustic system at hand, as visualized in Fig.\ref{fig:charSol} (blue circles). 

We investigate one of the unstable mode with $\omega_r/\omega_{p,1}=2.8$ and $\sigma/\omega_{p,1}=0.2$ (blue filled circle). Owing to the fact that the case at hand (ideal closed-open boundaries, $x_f = 0.4L$) is thus far discussed in detail, we could readily determine  that this mode is indeed an ITA mode using Fig. \ref{fig:FTFMap} or rather Fig. \ref{fig:PhiMap}. Nevertheless, it is crucial to realize that the proposed criterion that relates the angle of the velocity phasors across the flame should be valid for other general cases, where the flame position differs or the boundary conditions are non-ideal, say. In such cases, the convoluted process of solving the dispersion relation and producing the contour maps may be omitted. Instead, one could determine the `ITA-ness' of the given mode by comparing the corresponding FTF magnitude to the transitional value (Eq. \eqref{eq:FTF_transition}). To demonstrate this using the example at hand, we read off Fig.~\ref{fig:FTF_DTD} to obtain $|\mathcal{F}(\omega)|=1.8$ and $\phi_f(\omega)=-0.83\pi$. The FTF phase satisfies $\pi/2<|\phi_f(\omega)|<3\pi/2$ and yields a transitional FTF magnitude of $|\mathcal{F}_T(\omega)|=0.29$. With $|\mathcal{F}(\omega)|>|\mathcal{F}_T(\omega)|$, we come to the same conclusion that the mode at hand is an ITA mode. The phasor diagram in Fig.~\ref{fig:2Dphasor_DTD} also confirms this conclusion.
\begin{figure}[ht]
\centering
  \includegraphics[width=0.5\textwidth]{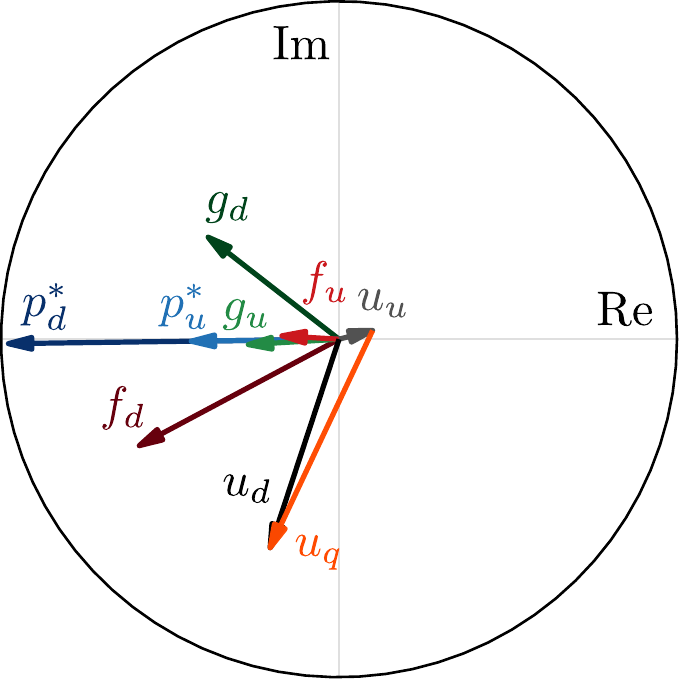}
\caption{2D phasor diagram of the mode with $\omega_r/\omega_{p,1}=2.8$ and $\sigma/\omega_{p,1}=0.2$. The partially out-of-phase $u_u$ and $u_d$ indicate an ITA mode. }
\label{fig:2Dphasor_DTD}
\end{figure}

\section{Continuity of  \texorpdfstring{$\criterionName$}{\criterionNameText} \label{apx:holomorphicProof}}

\cmnt{The real part of a holomorphic function is a harmonic function, which is a (continuously differentiable) analytic function. Consequently, it is possible to show that 
\begin{align}
    \criterionName = \tilde{u}_u\cdot \tilde{u}_d = \frac{1}{\left|u_u \right|\left|u_d \right|}\Re\left\{u_u^\dagger u_d\right\}
\end{align}
is a continuous (differentiable) function by showing that $u_u^\dagger u_d$ is holomorphic.} 

A complex function $f(z)\in \mathbb{C}$, which is dependent on the complex variable $z\in\mathbb{C}$
\begin{equation}
    f(z) = U(z) + iV(z), \quad z=x+iy
\end{equation}
is said to be holomorphic, if it satisfies the Cauchy-Riemann equations given in Eq.~\eqref{eq:CReq}
\begin{equation}
    U_x(x,y) = V_y(x,y) , \quad
    V_x(x,y) = -U_y(x,y), \label{eq:CReq}
\end{equation}
\cmnt{where the subscript $[\ ]_x , [\ ]_y$ denotes the differentiation with respect to $x$ and $y$, respectively.}

The product $u_u^\dagger u_d$ may be expanded to
\begin{align}
    u_u^\dagger u_d &= |u_u|^2\left(1+\theta \mathcal{F}\right) \\
    &= |u_u|^2\left( \underbrace{1+\theta |\mathcal{F}| e^{\phi_i}\cos(\phi_r)}_{U(\phi_r,\phi_i)}+i \underbrace{\left(-\theta |\mathcal{F}| e^{\phi_i}\sin(\phi_r)\right)}_{V(\phi_r,\phi_i)}\right).
\end{align}
where $\phi_f=\phi_r+i\phi_i$ is the complex phase of the general FTF model $\mathcal{F}=|\mathcal{F}| e^{-i\phi_f}$. 
Evaluating the Cauchy-Riemann equations yields 
\begin{align*}
		U_{\phi_r}(\phi_r,\phi_i) &= -\theta |\mathcal{F}| e^{\phi_i}\sin(\phi_r)\\ \nonumber
		V_{\phi_i}(\phi_r,\phi_i) &= -\theta |\mathcal{F}| e^{\phi_i}\sin(\phi_r)\\ \nonumber
		U_{\phi_i}(\phi_r,\phi_i) &= \theta |\mathcal{F}| e^{\phi_i}\cos(\phi_r)\\
		V_{\phi_r}(\phi_r,\phi_i) &= -\theta |\mathcal{F}| e^{\phi_i}\cos(\phi_r) \\[5pt]
\Rightarrow\quad
        U_{\phi_r}(\phi_r,\phi_i) &= V_{\phi_i}(\phi_r,\phi_i) \\
        U_{\phi_i}(\phi_r,\phi_i) &= -V_{\phi_r}(\phi_r,\phi_i). && \square
\end{align*}

\section{Comparison with \texorpdfstring{$\mu$}{mu} parameter sweep}

In the work of Emmert et al. \cite{EmmerBombe17}, a parameter $\mu$ was introduced to modulate \cmnt{the coupling strength between the acoustic and ITA feedback loops and thus} the dispersion relation obtained from the system matrix of a network model
\begin{align}
\mathcal{C}(\mu,\omega)\equiv& (1+\xi)(Z_d(\omega)- Z_u(\omega)\xi) \nonumber \\
 &+\mathcal{F}(\omega)\theta Z_d(\omega)(1+\xi\mu^2) + \mathcal{F}(\omega)\theta Z_u(\omega)\xi(1-\mu^2) =0. \label{eq:coupledEmmer}
\end{align}
With $\mu=1$, the dispersion relation represents the full physical system described by the network model, c.f. Eq.\eqref{eq:charEq}. However, if $\mu$ is reduced to 0, $\mathcal{C}$ is factorized into two partial equations, one describes a pure acoustic $\mathcal{C}_\text{AC}$ and the other describes a pure ITA system $\mathcal{C}_\text{ITA}$ \cite{EmmerBombe15}
\begin{equation}
\mathcal{C}(\mu=0,\omega) \equiv \underbrace{(Z_d(\omega)-Z_u(\omega)\xi)}_{\mathcal{C}_\text{AC}} \underbrace{(1+\xi+\mathcal{F}(\omega)\theta)}_{\mathcal{C}_\text{ITA}}. \label{eq:decoupEmmer}
\end{equation}
The idea of Emmert's method revolves around computing a numerical continuation path for each complex frequency as $\mu$ reduced gradually from 1 to 0, i.e. a parameter sweep. The algorithm given in Alg. \ref{alg:emmert} describes a natural parameter continuation method.
\begin{algorithm}
\caption{$\mu$ parameter sweep}\label{alg:emmert}
\begin{algorithmic}[1]
\Require $\omega_\text{init} \geq 0$
\Ensure $\mathcal{C}(\mu,\omega)= (1+\xi)(Z_d(\omega) - Z_u(\omega)\xi)+\mathcal{F}(\omega)\theta Z_d(\omega)(1+\xi\mu^2) + \mathcal{F}(\omega)\theta Z_u(\omega)\xi(1-\mu^2) $
\State $\mu \gets 1$
\State $\omega \gets \omega_\text{init}$
\State $\delta\mu \gets 1\text{E}-5$\Comment{$\delta\mu \ll 0$}
\While{$\mu >=0$}
    \State $\omega=\text{solve}(\mathcal{C}(\mu,\omega)=0)$
    \State $\mu \gets \mu-\delta\mu$ 
\EndWhile
\end{algorithmic}
\end{algorithm}
According to Emmert et al \cite{EmmerBombe17}, all modes that can be tracked to a corresponding pure acoustic mode at $\mu=0$ should be regarded as acoustic, while those that tracked to a pure ITA mode should be labeled as ITA. The continuation paths of a range of modes with the growth rate $\sigma/\omega_{p,1}=0.05$ and real valued frequencies $\omega_r/\omega_{p,1}\in(0.25,3.25)$ are plotted in the complex plane shown in Fig. \ref{fig:emmertTrack}. To distinguish the modes identified with the $\mu$-sweep method from those with the phasor based method proposed in this work, we name the former \emph{$\mu$-acoustic} or \emph{$\mu$-ITA} modes, while the latter simply acoustic or ITA modes to preserve consistency with the rest of the paper. 
\begin{figure}[ht]
\vspace{10mm}
\centering
\includegraphics[width=0.6\textwidth]{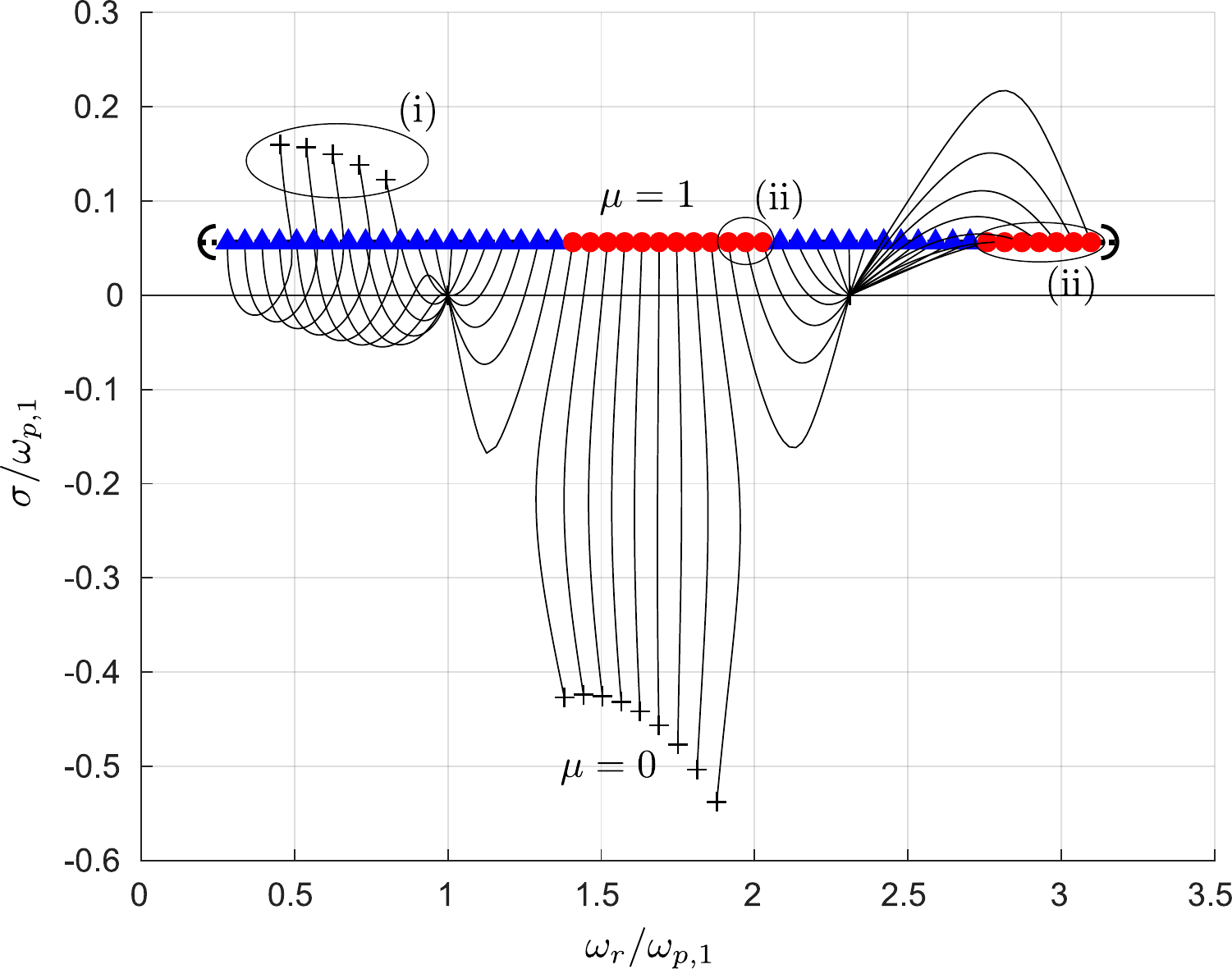}
\caption{Continuation paths of thermoacoustic modes with $\omega_r/\omega_{p,1}\in(0.25,3.25)$ and $\sigma/\omega_{p,1}=0.05$, as $\mu$ is reduced from 1 to 0. The colored markers represent the thermoacoustic modes of the full system $\mu=1$, while the crosses represent the pure acoustic or ITA modes, to which the thermoacoustic modes are tracked at $\mu=0$. The results from these two methods show rough agreements. Mismatching results are indicated with the ovals: (i) acoustic modes that tracked to pure ITA modes, (ii) ITA modes that tracked to pure acoustic modes. }
\label{fig:emmertTrack}
\end{figure}
The results from the $\mu$ parameter sweep indicate rough agreements with those from the $\criterionName$ method. To evaluate the general conformity between these two methods, the $\mu$ parameter sweep is performed on all thermoacoustic modes of interest in the range identical to Fig. \ref{fig:PhiMap}. The results are visualized in Fig. \ref{fig:emmertFreq}.
\begin{figure}[ht]
\centering
\includegraphics[width=0.6\textwidth]{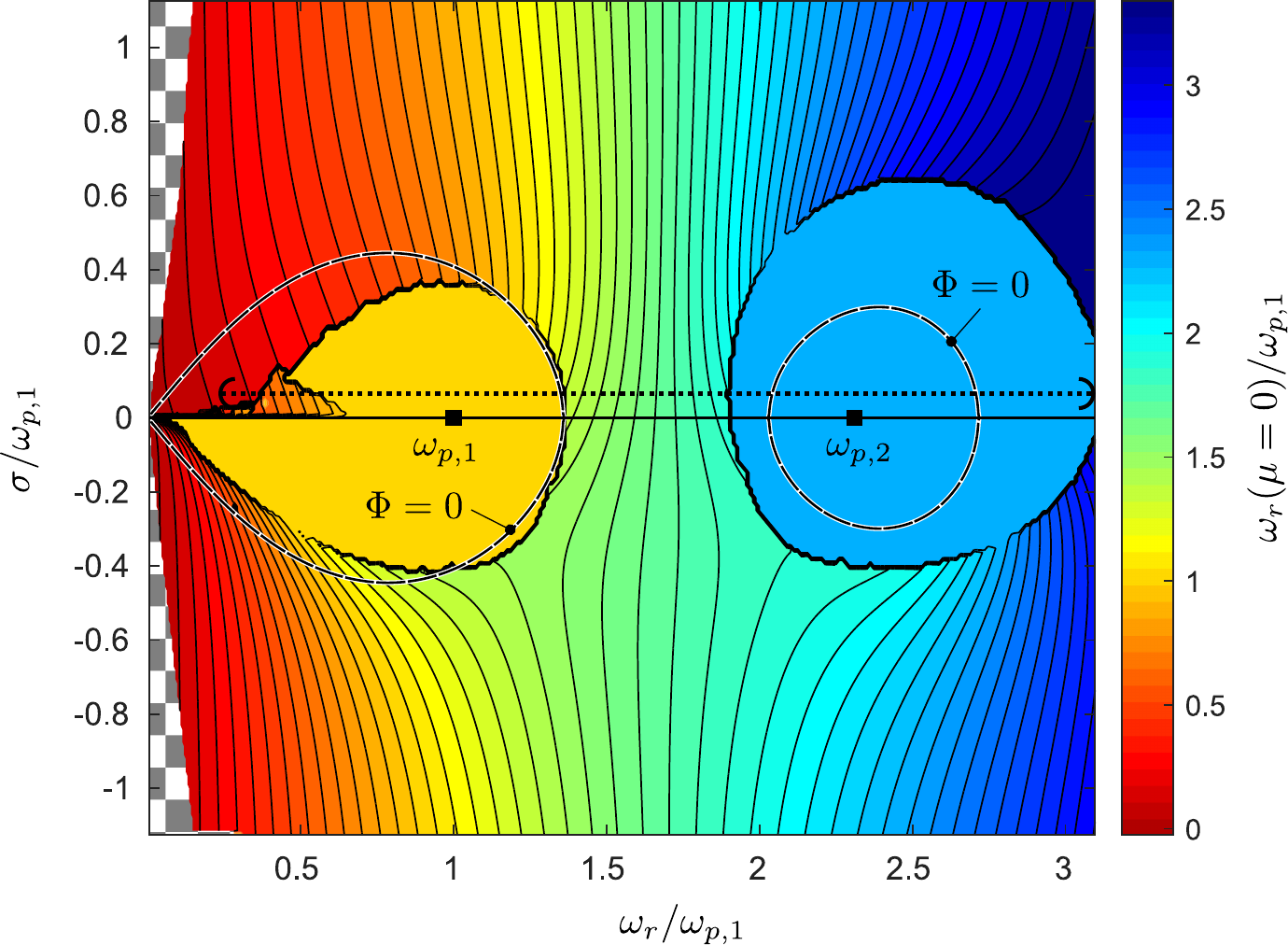}
\caption{Contour plot depicting the destination frequency of each thermoacoustic mode within the range of $\omega_r/\omega_{p,1}\in(0,3.1]$ and $\sigma/\omega_{p,1}\in[-1.2,1.2]$, as $\mu$ reduced from 1 to 0. The $\mu$-acoustic modes constitutes the homogeneous yellow and blue (irregular) patches around $\omega_{p,i}$. Beyond the patches are the $\mu$-ITA modes. For comparison purpose, the black thickened curves from Fig. \ref{fig:PhiMap} (now dashed to improve visibility), which 
marks the acoustic to ITA transition is included. The corresponding modes that are shown in Fig. \ref{fig:emmertTrack} are indicated with the dotted horizontal line.}
\label{fig:emmertFreq}
\end{figure}

\begin{figure}[ht]
\centering
\includegraphics[width=0.6\textwidth]{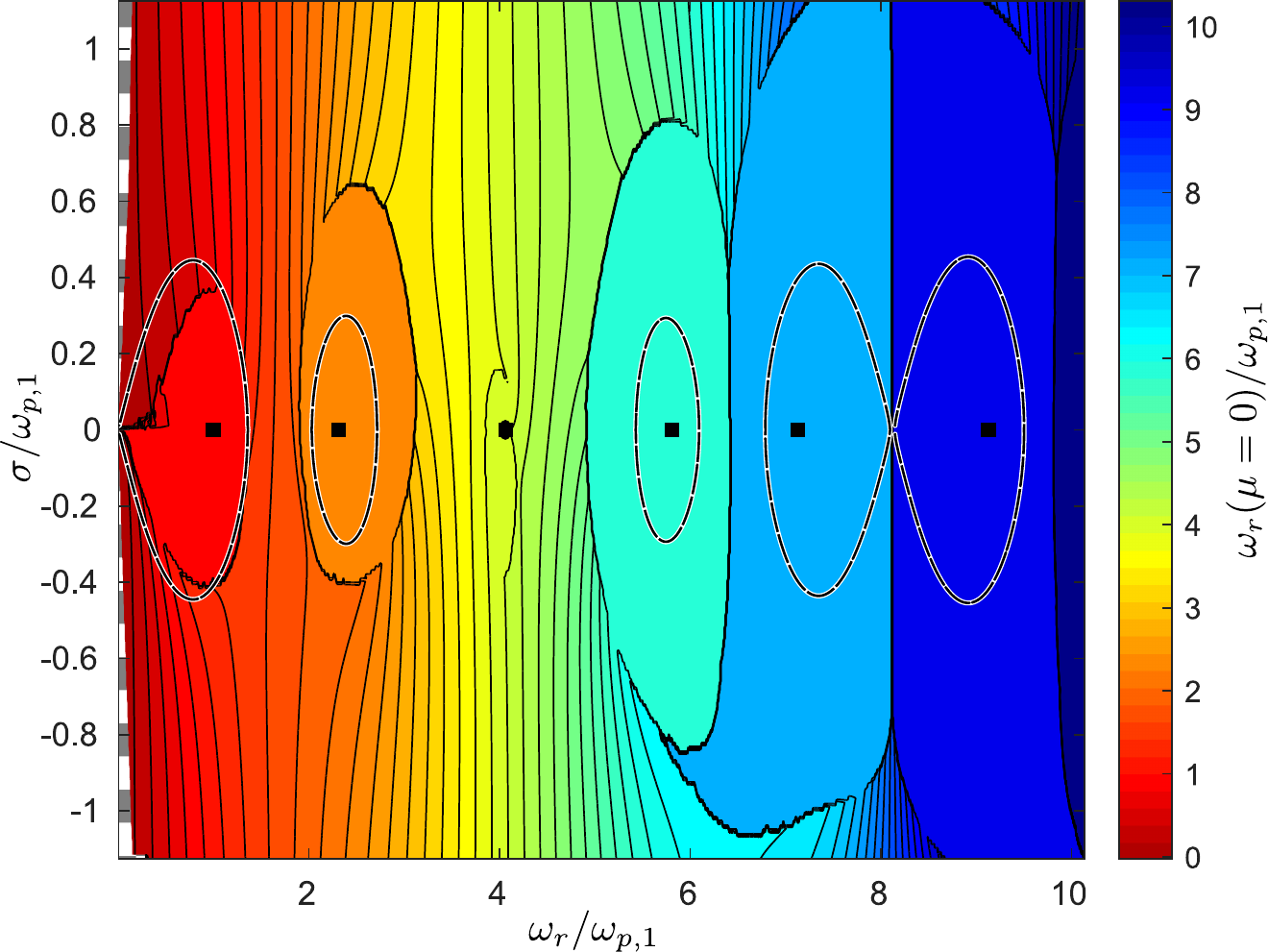}
\caption{Contour plot depicting the destination frequency of each thermoacoustic mode within the range of $\omega_r/\omega_{p,1}\in(0,3.1]$ and $\sigma/\omega_{p,1}\in[-1.2,1.2]$, as $\mu$ reduced from 1 to 0. The irregularity is more significant at higher frequencies $\omega_r/\omega_{p,1}>5$, such that even highly damped and growing modes $|\sigma|/\omega_{p,1}>1$ are categorized as acoustic. }
\label{fig:emmertFreq_large}
\end{figure}

The $\mu$-acoustic region around the first passive acoustic mode $\omega_{p,1}$ agrees roughly with the acoustic region identifiable in Fig. \ref{fig:PhiMap}. However, the same cannot be said for that around the second pure acoustic mode $\omega_{p,2}$. Despite sharing a similar feature of occupying a defined region around $\omega_{p,2}$, the $\mu$-acoustic region is significantly larger. Another obvious disparity is the shape irregularity. Disregarding the rugged border, which is excusable due to the numerical nature of the algorithm, a reason cannot be deduced as to why the shape of the $\mu$-acoustic region is not symmetric to the real axis, when the symmetric dispersion relations Eqs. \eqref{eq:coupledEmmer} and \eqref{eq:decoupEmmer} are. This irregularity increases at higher frequencies as shown in Fig. \ref{fig:emmertFreq_large}.  
The expansion of the $\mu$-acoustic region into regions of high growth and decay rates $|\sigma|/\omega_{p,1}>1$ might not be physical. As discussed, the outgoing waves and incoming waves are highly localized at the flame at these growth and decay rates, \cmnt{rendering the upstream and downstream boundaries irrelevant}. Hence, these modes should be more physically similar to the pure ITA modes than the pure acoustic modes. Further details on the growing $\mu$-acoustic regions are not developed in this work. 

In any case, categorization methods based on a parameter sweep suffer from multiple drawbacks. Among others, their computations are relatively costly due to the iterative nature. Besides, strong parameter variations would result in modifications to the original system of interest, thereby defeating the purpose of mode categorization. Moreover, parameter sweeps are difficult to implement if not impossible in experimental environments. Not to mention that categorization by parameter sweeps are not related to  physical characteristics in terms of pressure gradient or sign of velocity across the flame. In comparison, the categorization framework proposed in this work requires only the distribution of acoustic variables within the thermoacoustic system, and more importantly the frequency and FTF of the system, which are readily available in simulations as well as in experiments that analyze thermoacoustic systems.

\section{Mathematics of the oscillation velocities \label{apx:oscillation}}

\begin{figure}[ht]
\centering
\begin{subfigure}{.3\textwidth}
  \centering
  \includegraphics[width=\linewidth]{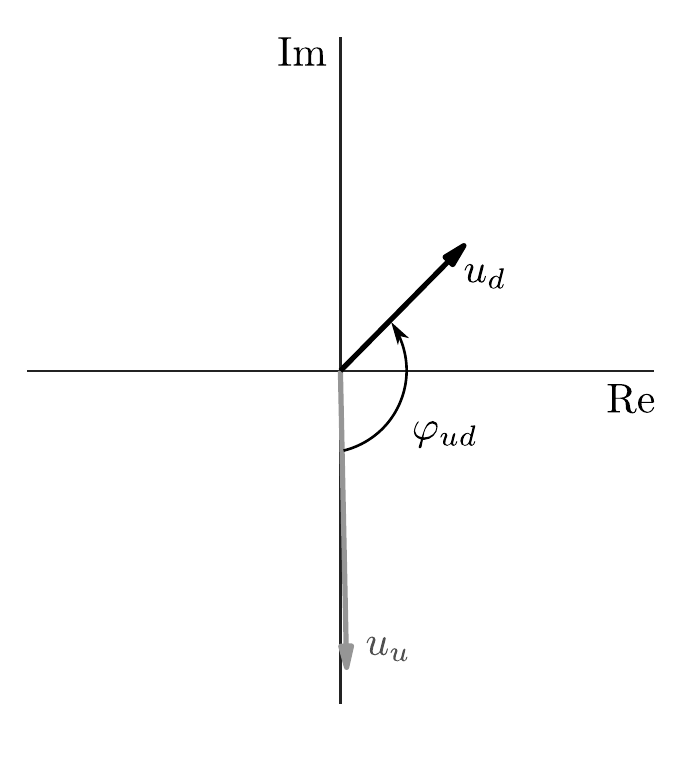}
  \caption{}
\end{subfigure}%
\hspace{.05\textwidth}
\begin{subfigure}{.55\textwidth}
  \centering
  \includegraphics[width=\linewidth]{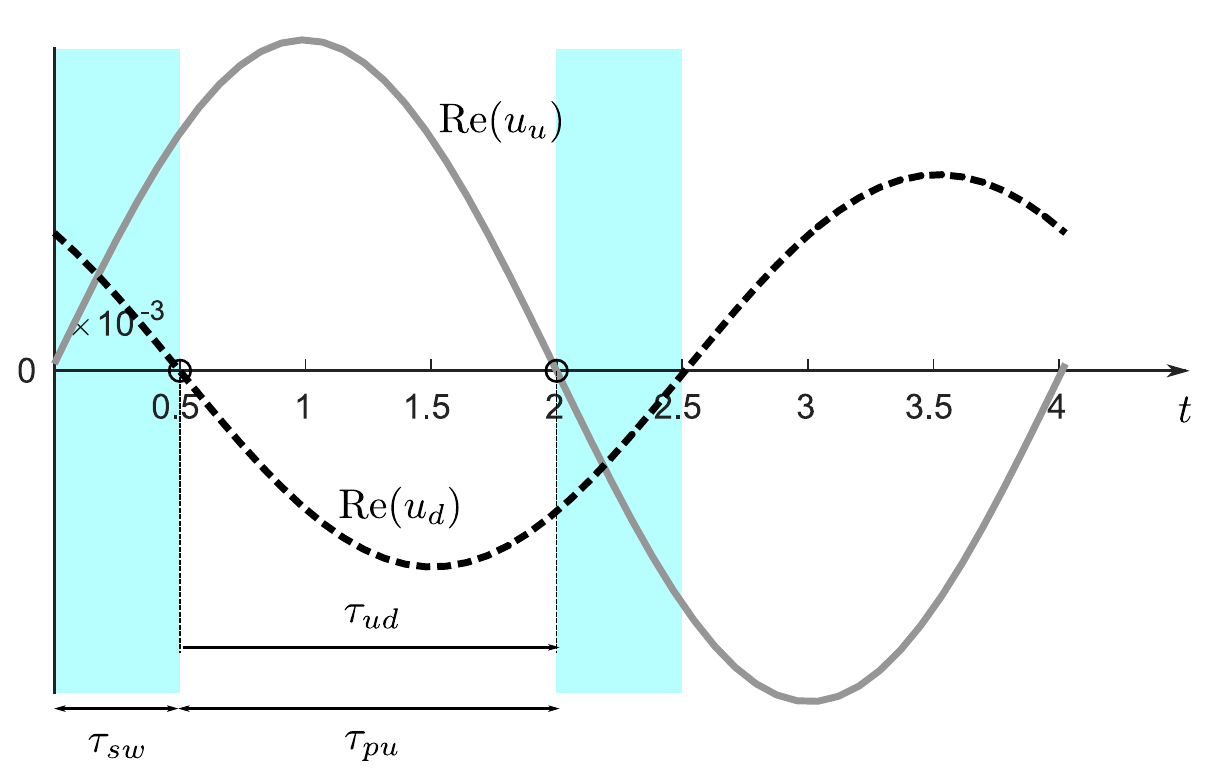}
  \caption{}
\end{subfigure}
\caption{(a) Phasor diagram showing an ITA mode with imperfect direction reversal of the velocity phasors. (b) Plot showing the displacement rates on the upstream (grey solid curve) and downstream (black dashed curve) of the flame with respect to time. The time delay between the zero displacement rates on the upstream and that of the downstream $\tau_{ud}$ corresponds to the relative phase between the velocity phasors $\varphi_{ud}=\omega_r\tau_{ud}$. It is clear that the duration of a pulsating oscillation $\tau_\text{pu}$, i.e. oscillation rates with opposite signs, equates $\tau_{ud}$. It is implied that the medium at the flame pulsates more frequently than it swings for an ITA mode.}
\label{fig:u-t}
\end{figure}

Fig. \ref{fig:u-t} shows the medium velocities on the flame upstream and downstream with respect to time for an ITA mode. Note that the amplitudes of the plots are normalized with the factor $e^{\sigma t}$ to eliminate the amplitude growth of the oscillations and thereby ease the visualization. The pulsating oscillation (white region) is observed more frequently in a single cycle than the swinging oscillation (cyan region).
The proportion of time, in which the medium swings $r_\text{sw}$ or pulsates $r_\text{pu}$, is given by
\begin{equation}
r_\text{pu} = \frac{\varphi_{ud}}{\pi}; \quad r_\text{sw}=1-r_\text{pu}
\end{equation}
where $\varphi_{ud} = \omega_r\tau_{ud}$ and $\tau_{ud}=\tau_\text{pu}$. It is evident that if $u_u,u_d$ are partially anti-aligned, then $r_\text{pu}>r_\text{sw}\Leftrightarrow r_\text{pu}>0.5$ 
, signifying a dominant pulsating oscillation -- an ITA mode. 

The same $\criterionName$ criterion may be derived by averaging the product of the displacement rates over a  multiple of oscillation cycles $T = 2k\pi/\omega_r$  
\begin{align}
\criterionName &= \frac{2}{T}\int_0^T \Re(\tilde{u}_u)\Re(\tilde{u}_d) dt\\
&=\underbrace{\frac{2}{T}\cos\varphi_{ud}\int_0^T \cos^2(\omega_r t) \ dt}_{\cos\varphi_{ud}} -  \underbrace{\frac{2}{T}\sin\varphi_{ud}\int_0^T \cos(2\omega_r t) \ dt}_{0}\\
&= \tilde{u}_u \cdot \tilde{u}_d. 
\end{align}